\documentclass[twocolumn]{aastex63}
\usepackage{amsmath}
\usepackage{multirow}
\usepackage{xcolor}

\accepted{February 27th, 2020}
\submitjournal{ApJ}

\shorttitle{Depleted inner disks in transition disks}
\shortauthors{Francis and van der Marel}
\graphicspath{{./}{figures/}}
\begin{document}

%\title{Interior Transient/Small(?) Dust Rings in resolved ALMA observations of 38 Transition Disks}

\title{Dust depleted inner disks in a large sample of transition disks through long-baseline ALMA observations}

\author[0000-0001-8822-6327]{Logan Francis}
\affiliation{Department of Physics and Astronomy, University of Victoria \\
3800 Finnerty Road, Elliot Building \\
Victoria, BC, V8P 5C2, Canada}
\affiliation{NRC Herzberg Astronomy and Astrophysics \\
5071 West Saanich Road \\
Victoria, BC, V9E 2E7, Canada}
\affiliation{loganfrancis3@uvic.ca}

\author[0000-0003-2458-9756]{Nienke van der Marel}
\affiliation{Department of Physics and Astronomy, University of Victoria \\
3800 Finnerty Road, Elliot Building \\
Victoria, BC, V8P 5C2, Canada}
\affiliation{NRC Herzberg Astronomy and Astrophysics \\
5071 West Saanich Road \\
Victoria, BC, V9E 2E7, Canada}
\affiliation{Banting fellow}
\affiliation{astro@nienkevandermarel.com}

%\collaboration{1}{(AAS Journals Data Scientists collaboration)}



\begin{abstract}   
% 250 word limit!

% 251 words including numbers, should be fine if numbers don't count.
Transition disks with large inner dust cavities are thought to host massive companions. However, the disk structure inside the companion orbit and how material flows toward an actively accreting star remain unclear. We present a high resolution continuum study of inner disks in the cavities of 38 transition disks. Measurements of the dust mass from archival Atacama Large Millimeter/Submillimeter Array observations are combined with stellar properties and spectral energy distributions to assemble a detailed picture of the inner disk. An inner dust disk is detected in 18 of 38 disks in our sample. Of the 14 resolved disks, 8 are significantly misaligned with the outer disk. The near-infrared excess is uncorrelated with the mm dust mass of the inner disk. The size-luminosity correlation known for protoplanetary disks is recovered for the inner disks as well, consistent with radial drift. The inner disks are depleted in dust relative to the outer disk and their dust mass is uncorrelated with the accretion rates.  This is interpreted as the result of radial drift and trapping by planets in a low $\alpha$ ($\sim 10^{-3}$) disk, or a failure of the $\alpha$-disk model to describe angular momentum transport and accretion. The only disk in our sample with confirmed planets in the gap, PDS 70, has an inner disk with a significantly larger radius and lower inferred gas-to-dust ratio than other disks in the sample. We hypothesize that these inner disk properties and  the detection of planets are due to the gap having only been opened recently by young, actively accreting planets.

\end{abstract}



\section{Introduction}
Protoplanetary disks are circumstellar reservoirs of gas and dust found surrounding young stars, and are the presumed birthplace of planets. Transition disks are a class of protoplanetary disks with large cavities in their dust distributions seen directly in millimetre observations  \citep{andrews2011} or implied from a deficit of infrared emission in their spectral energy distribution (SED) \citep{Espaillat2014}. The dust cavities in transition disks can be formed by embedded companions  \citep[e.g.][]{LinPapaloizou1979} or dead zones \citep[e.g.][]{Regaly2012}, which can create local enhancements in the gas pressure where dust grains are trapped and grow to mm sizes \citep{Pinilla2012b}. This trapping is necessary to explain the observed emission at millimetre wavelengths, as otherwise drag forces from the gas acting  on the large (mm-cm) size grains would cause them to rapidly drift into the star \citep{Whipple1972, Weidenschilling1977}. Trapping by a companion is strongly favoured due to the detection of deep gas cavities within dust cavities in several transition disks by the Atacama Large Millimeter/Submillimeter Array (ALMA) \citep[e.g.][]{vanderMarel2016-isot,Dong2017}. The only robust detection of a planet embedded inside a transition disk cavity so far, however, is PDS70b \citep{Keppler2018,Haffert2019}. Other planet candidates inside cavities have been claimed \citep[e.g.][]{Quanz2013,Sallum2015}, but later discarded as imaging artefacts \citep[e.g.][]{Rameau2017,Currie2019}.

Several lines of evidence indicate that dust trapping does not produce transition disk cavities completely devoid of dust, and that an \emph{inner disk} exists close to the star. The presence of a significant near-infrared (NIR) excess over the stellar blackbody in some objects suggests an inner ring of micron-sized warm dust extending to the sublimation radius separated from the outer disk \citep{Espaillat2007}, which has been used to define the "pre-transition disk" (PTD) sub-class. This distinction between pre-transition disk and transition disk has been suggested as an evolutionary sequence, where the inner disk is eventually cleared by accretion onto the star. However, the modelling of the inner disk responsible for the NIR excess is subject to several uncertainties, in particular, the opacity and the assumed 3D geometry, which makes it difficult to constrain the extent of the dust distribution. 

The high accretion rates in transition disks, comparable to rates in 'full' protoplanetary disks, require the presence of a gas-rich inner disk \citep[e.g.][]{Manara2014}, or flow through the cavity at free-fall speeds if the gas is also depleted \citep{Rosenfeld2014}. ALMA observations of CO isotopologues indicate that the gas surface density drops by several orders of magnitude within transition disk dust cavities, however, these studies were limited to a resolution of $\sim 30$ au and thus unable to detect the gas structure in the inner part of the disk \citep[e.g.][]{vanderMarel2016-isot}. Detections of rovibrational CO line emission, which is only excited at high temperatures, also indicates potential high gas surface densities close to the star \citep[e.g.][]{Pontoppidan2008,BanzattiPontoppidan2015}. Models of viscous (``$\alpha$'') accretion disks \citep{Shakura1973} predict a tight correlation between the total disk mass and accretion rate \citep{Hartmann1998}, which has recently been found observationally in protoplanetary disks \citep{Manara2016, Sanchis2019}. This favors a gas-rich inner disk, however, the validity of $\alpha$-disk models has been called into question by observational and theoretical concerns over the degree of turbulence present in protoplanetary disks \citep{Flaherty2017,Flaherty2018,Bai2016}.

A \emph{misaligned} inner disk has been invoked to explain several phenomena in transition disks. Scattered light observations show shadowing of the outer disk in several objects (e.g. DoAr~44, \citet{Casassus2018}, RXJ1604.3~-~2130, \citet{Pinilla2018}) which are well described by a misaligned or warped inner disk. Observations of the gas motion as traced by CO also display velocity patterns consistent with a misaligned inner gas disk (e.g. AA Tau, \citet{Loomis2017}). Finally, a misaligned inner disk has been used to explain the aperiodic/quasi-period dips seen in the light curves of dipper stars, some of which are known to host transition disks (AA~Tau, \citet{Bouvier1999, Bouvier2007}, RXJ1604.3~-~2130 \citet{Sicilia-Aguilar2019}).
    
While an inner disk has been inferred from the previous findings indirectly, a properly quantified study of inner disks in a large sample of transition disks is still lacking. The mm-dust content of an inner disk can be measured by long-baseline observations with ALMA. For nearby transition disks, ALMA is capable of resolving dust emission on scales as small as 5 au (equivalent to a 0.033\arcsec~ beam at 150 pc), and several inner mm-dust disks have previously been detected, e.g. HD~142527, DM Tau and SR24S \citep{Fukagawa2013,Kudo2018,Pinilla2019}. 

In this work, we present an ALMA continuum study of inner disks in a sample of 38 transition disks. We combine archival ALMA observations with a range of sensitivities and spatial resolutions at typically $0.05-0.1\arcsec$ or $7-15$ au resolution with measurements of the SED-derived near-infrared (NIR) excess and accretion rates to investigate the properties of the inner disks.

The paper is organized as follows: In Section \ref{sec:obs}, we present our sample selection (\ref{ssec:sample_sel}), the ALMA data reduction process (\ref{ssec:alma_dr}), the collection of the stellar properties and SED photometry (\ref{ssec:stellar_props}), and the calculation of the NIR excess (\ref{ssec:nir_excess_calc}). In Section \ref{sec:analysis}, we outline our methods of fitting the inner disk emission (\ref{ssec:inner_fits}), determining the inner disk dust mass (\ref{ssec:inner_dust_mass}), and measuring global dust surface density profiles (\ref{ssec:outer_disk}). In Section \ref{sec:discussion}, we discuss misalignments between the inner and outer disk (\ref{ssec:misalignment_disc}), correlations between mm-dust mass and NIR excess or accretion rates (\ref{ssec:inner_nir},\ref{ssec:inner_dust}), the size of the inner dust disk compared with the size-luminosity relation and the water snowline (\ref{ssec:innerdisksize}), the gas content and angular momentum transport through the inner disk (\ref{ssec:inner_gas}), and the consequences for the presence of companions (\ref{ssec:companions}). A summary of the paper and our conclusions are given in Section \ref{sec:conclusions}. Plots of the SEDs for our sample, the determination of the inclination and position angle of some outer disks, and the coordinates of the detected inner disks are given in the Appendix.

\label{sec:intro}
 
%\newpage
\section{Sample and Observations}
\label{sec:obs}

\subsection{Sample selection}
\label{ssec:sample_sel}

Our sample consists of 38  transition disks with large inner dust cavities ($>$25 au, except TW Hya), for which spatially resolved ALMA archival data exist and for which the spatial resolution is sufficient to fully resolve the dust cavity. If fully resolved, the millimeter flux of the inner disk (when detected) can be measured directly and its size can be constrained. For the majority of disks, Band 6 (1.3 mm or 230 GHz) or Band 7 (0.87 mm or 345 GHz) data at high resolution is available, but for four targets (HD~135344B, T~Cha, SR21, SR24S) data in Band 3 (3.0 mm or 100 GHz) actually had higher spatial resolution and those were chosen for our analysis instead. The full list of targets and their respective observing wavelength and resolution is given in Table \ref{tab:disk_sample}. About half of our datasets have been published already; references are provided in the same table. 

\begin{deluxetable*}{cchcccccc} % ALMA band hidden!
\label{tab:disk_sample}
\tablecaption{Sample and deconvolved image properties.}
\tablehead{\colhead{Name} & \colhead{ALMA Project IDs} & \nocolhead{ALMA Band} & \colhead{Frequency} & \colhead{Beam} & \colhead{RMS} & \colhead{Briggs robust} & \colhead{Ref.}\\ \colhead{ } & \colhead{ } & \colhead{ } & \colhead{(GHz)} & \colhead{(\arcsec)} & \colhead{(mJy bm$^{-1}$)} & \colhead{ } & \colhead{}}
\startdata
AATau & 2013.1.01070.S, 2016.1.01205.S & 6 & 241.1495 & 0.07 x 0.04 & 0.04 & {-2.0} & 1,2\\
ABAur & 2015.1.00889.S & 6 & 232.7003 & 0.06 x 0.03 & 0.02 & 2.0 & 3\\
CIDA9 & 2016.1.01164.S & 6 & 217.9947 & 0.14 x 0.11 & 0.07 & 0.5 & 4\\
CQTau & 2016.A.00026.S, 2017.1.01404.S & 6 & 224.9946 & 0.07 x 0.05 & 0.02 & 0.5 & 5\\
CSCha & 2017.1.00969.S & 7 & 341.1324 & 0.09 x 0.06 & 0.04 & 0.5 & 6\\
DMTau & 2013.1.00498.S, 2017.1.01460.S & 6 & 224.8051 & 0.04 x 0.03 & 0.01 & 2.0 & 7\\
DoAr44 & 2012.1.00158.S & 7 & 336.0885 & 0.22 x 0.19 & 0.33 & {-2.0} & 8\\
GGTau AA/Ab & 2012.1.00129.S, 2015.1.00224.S & 7 & 336.8155 & 0.26 x 0.16 & 1.23 & 0.5 & 9,10\\
GMAur & 2015.1.01207.S, 2017.1.01151.S & 6 & 243.1163 & 0.04 x 0.02 & 0.01 & 0.5 & 11\\
HD100453 & 2015.1.00192.S, 2017.1.01424.S & 7 & 281.2888 & 0.03 x 0.02 & 0.03 & {-2.0} & 12\\
HD100546 & 2016.1.00344.S & 6 & 224.9983 & 0.04 x 0.02 & 0.03 & {-2.0} & 13\\
HD135344B & 2017.1.00884.S & 3 & 109.1009 & 0.11 x 0.07 & 0.03 & 2.0 & 14\\
HD142527 & 2012.1.00631.S & 7 & 321.7245 & 0.12 x 0.09 & 0.13 & -2.0 & - \\
HD169142 & 2016.1.00344.S & 6 & 225.0001 & 0.05 x 0.03 & 0.01 & 0.5 & 15\\
HD34282 & 2015.1.00192.S, 2017.1.01578.S & 6 & 225.5628 & 0.06 x 0.05 & 0.02 & 0.5 & 16\\
HD97048 & 2016.1.00826.S & 7 & 338.0715 & 0.06 x 0.03 & 0.12 & {-2.0} & 17\\
HPCha & 2017.1.01460.S & 6 & 224.8087 & 0.05 x 0.03 & 0.02 & 0.5 & 18\\
IPTau & 2016.1.01164.S & 6 & 225.4950 & 0.12 x 0.09 & 0.07 & {-2.0} & 4\\
IRS48 & 2013.1.00100.S & 7 & 343.0859 & 0.13 x 0.10 & 0.24 & {-2.0} & 19\\
J1604.3-2130 & 2015.1.00888.S & 7 & 349.7554 & 0.13 x 0.12 & 0.14 & {-2.0} & 20\\
LkCa15 & 2015.1.00678.S & 7 & 284.1257 & 0.18 x 0.12 & 0.16 & {-2.0} & 21\\
MHO2 & 2013.1.00498.S & 6 & 224.6686 & 0.20 x 0.14 & 0.40 & {-2.0} & 22\\
MWC 758 & 2017.1.00492.S & 7 & 343.5038 & 0.05 x 0.04 & 0.04 & 0.5 & 23\\
PDS70 & 2015.1.00888.S, 2017.A.00006.S & 7 & 350.5927 & 0.07 x 0.05 & 0.03 & 0.5 & 24\\
PDS99 & 2015.1.01301.S & 6 & 226.4268 & 0.24 x 0.15 & 0.14 & {-2.0} & 25\\
RXJ1842.9-3532 & 2015.1.01083.S & 7 & 343.5127 & 0.16 x 0.12 & 0.12 & {-2.0} & 26\\
RXJ1852.3-3700 & 2015.1.01083.S & 7 & 343.5125 & 0.15 x 0.12 & 0.17 & {-2.0} & 27\\
RYLup & 2017.1.00449.S & 6 & 220.7937 & 0.12 x 0.11 & 0.06 & {-2.0} & 28\\
RYTau & 2013.1.00498.S, 2017.1.01460.S & 6 & 224.7880 & 0.04 x 0.02 & 0.07 & 0.5 & 29\\
SR21 & 2017.1.00884.S & 3 & 108.0000 & 0.10 x 0.07 & 0.02 & 2.0 & 30\\
SR24S & 2017.1.00884.S & 3 & 108.0000 & 0.07 x 0.06 & 0.04 & 0.5 & 31\\
Sz91 & 2012.1.00761.S & 7 & 349.4095 & 0.15 x 0.12 & 0.04 & 0.5 & 32\\
Tcha & 2015.1.00979.S & 3 & 97.4942 & 0.08 x 0.04 & 0.03 & {-2.0} & 33\\
TWHya & 2011.1.00399, 2013.1.00198.S, 2015.1.00686.S & 7 & 345.8636 & 0.02 x 0.02 & 0.04 & 0.0 & 34\\
UXTauA & 2015.1.00888.S & 7 & 349.7559 & 0.15 x 0.12 & 0.33 & {-2.0} & 35\\
V1247Ori & 2015.1.00986.S & 7 & 351.4541 & 0.03 x 0.03 & 0.06 & 0.5 & 36\\
V4046Sgr & 2017.1.01167.S & 6 & 238.8056 & 0.06 x 0.04 & 0.05 & 0.5 & 37\\
WSB60 & 2016.1.01042.S & 6 & 224.6866 & 0.11 x 0.08 & 0.10 & {-2.0} &-
\enddata
\tablecomments{
{\bf Refs.} 1) \citet{Loomis2017}; % AA Tau
2) Loomis et al. in prep. % AA Tau
3) \citet{Tang2017}; % AB Aur
4) \citet{Long2018}; % CIDA 9
5) \citet{Ubeira2019}; % CQ Tau
6) Kurtovic et al. in prep.; % CSCha
7) \citet{Kudo2018}; % DM Tau
8) \citet{vanderMarel2016-isot}; % DoAr44
9) \citet{Dutrey2016}; % GG Tau AA/Ab
10) \citet{Phuong2020}; % GG Tau AA/Ab 
11) \citet{Huang2020}; % GM Aur
12) \citet{Rosotti2019}; % HD100453
13) \citet{Perez2020}; % HD100546
14) \citet{Cazzoletti2018}; % HD135344B
% -  HD 142527, no publications or in prep
15)  \citet{PerezS2019}; % HD169142
16)  de Boer et al. in prep.; % HD34282
17)  van der Plas et al. subm.; % HD97048
18) Konishi et al. in prep.; % HP Cha
%4) \citet{Long2018}; % IPTau
19) van der Marel et al. in prep.; % IRS48
20) \citet{Mayama2018}; % J1604.3-2130
21) \citet{Qi2019}; % LkCa15  
22) \citet{Pinilla2018} % MHO2    
23) \citet{Dong2018}; % MWC 758/HD36112 
24) \citet{Keppler2019}; % PDS70   
25) Hashimoto et al. in prep.; % PDS99   
26) Morino \& Fukagawa in prep.; % RXJ1842.9-3532
27) \citet{Villenave2019}; % RXJ1852.3-3700  
28) van der Marel et al. in prep.; % RYLup
29) Konishi et al. in prep.; % RYTau   
30) Muro-Arena et al. subm.; % SR21
31) \citet{Pinilla2019}; % SR24S
32) \citet{Tsukagoshi2019}; % Sz91    
33) \citet{Hendler2018}; % Tcha    
34) \citet{Andrews2016}; % TWHya
35) Akiyama et al. in prep.; % UXTauA  
36) \citet{Kraus2017}; % V1247Ori    
37) Perez et al. in prep. % V4046Sgr    
% -  WSB60/ISO Oph 196, no publications or in prep
}
\end{deluxetable*}

The final sample consists of transition disks in nearby star forming regions such as Taurus, Chamaeleon, Ophiuchus, Lupus, Upper Sco, Corona Australis, TW Hydra and a number of isolated objects (all Herbig stars). Most targets are located within 200 pc with the exception of V1247~Ori and HD~34282, at 400 and 312 pc distance, respectively. 

Biases in the sample towards large and bright objects likely exist: several disks were observed with ALMA at high spatial resolution as targeted observations on the disk structure, as the presence of their cavity was already known from previous (sub)millimeter observations and/or a dip in the infrared part of their Spectral Energy Distribution (SED). Some transition disks have been discovered serendipitously as part of complete disk surveys of nearby star forming regions  \citep[e.g.][]{Ansdell2016,Pascucci2016,Cox2017,Long2018} at lower spatial resolution ($\sim$0.12-0.3"). However, for many disks identified in those surveys the resolution was insufficient to fully resolve the cavity and they were thus not included. This applies in particular to the new transition disks in Lupus \citep{vanderMarel2018} and Chamaeleon \citep{Pascucci2016}. The sample is thus not a \emph{complete} sample of all transition disks with large cavities within 200 pc, but representative across a wide range of disk and stellar properties. Spectral types range between A0 and M6 with an increased occurrence of early type objects. 

For Herbig Group I disks (Herbig stars with high far infrared excess, considered the Herbig transition disks) within 400 pc the completeness of the sample can be readily estimated: 9 out of 12 Group I Herbig stars as identified by \citet{Garufi2017} have been spatially resolved with ALMA and are included in our sample. The only exceptions are HD~141569, which is a much more evolved object \citep{White2016}, and HD~179218 and HD~139614, which do not have high-resolution ALMA data yet. 

For T Tauri stars completeness is much harder to determine, as these are taken from both targeted studies and disk surveys, and it remains unclear if all transition disks are known within these regions. The disk surveys of nearby star forming regions (dominated by T Tauri stars) are both resolution, sample and sensitivity limited, but with different limits for each region. For example, Chamaeleon was observed at a modest $\sim$0.6" resolution \citep{Pascucci2016}, leaving all but the largest ($>$50 au) cavities unresolved, 
whereas Taurus and Ophiuchus were observed at $\sim$0.12" \citep{Long2018,Cieza2019}, but the full Class II population of Taurus has not been covered yet. The Lupus disk survey was complete \citep{Ansdell2016,Ansdell2018} but at a resolution of 0.25-0.3", and even though 11 transition disks were identified \citep{vanderMarel2018}, only 2 were followed up in high enough resolution to resolve the inner disk. However, these disk surveys do indicate that transition disks with large cavities are generally bright \citep{OwenClarke2012,vanderMarel2018}, so even if our sample is incomplete, the disks selected here likely cover the majority of the transition disks with large cavities within the Solar Neighborhood.

\subsection{ALMA data reduction}
\label{ssec:alma_dr}

The archival ALMA data collected for each disk was chosen in order to best resolve the dust cavity and inner disk (if present). All data were calibrated using the ALMA pipeline reduction scripts provided with the raw data from the ALMA archive, except TW Hya, for which the reduced image from \citet{Andrews2016} was used.  

Images created from long baseline data alone can suffer from negative bowls around bright structures, which can potentially lower the flux of the inner disk. To avoid this and maximize the sensitivity to the inner disk, we verified that each ALMA project contained sufficient short baselines to recover the largest scales in the disk. Where necessary, long baseline observations were combined with more compact configuration ALMA data from other projects in order to recover the largest scales in each disk. The only image missing compact configuration data is AB Aur, as no Band 6 observations in a compact configuration have been taken, and the outer disk is consequently resolved out.

\begin{figure*}[htb]
    \centering
    \includegraphics[scale=1.0]{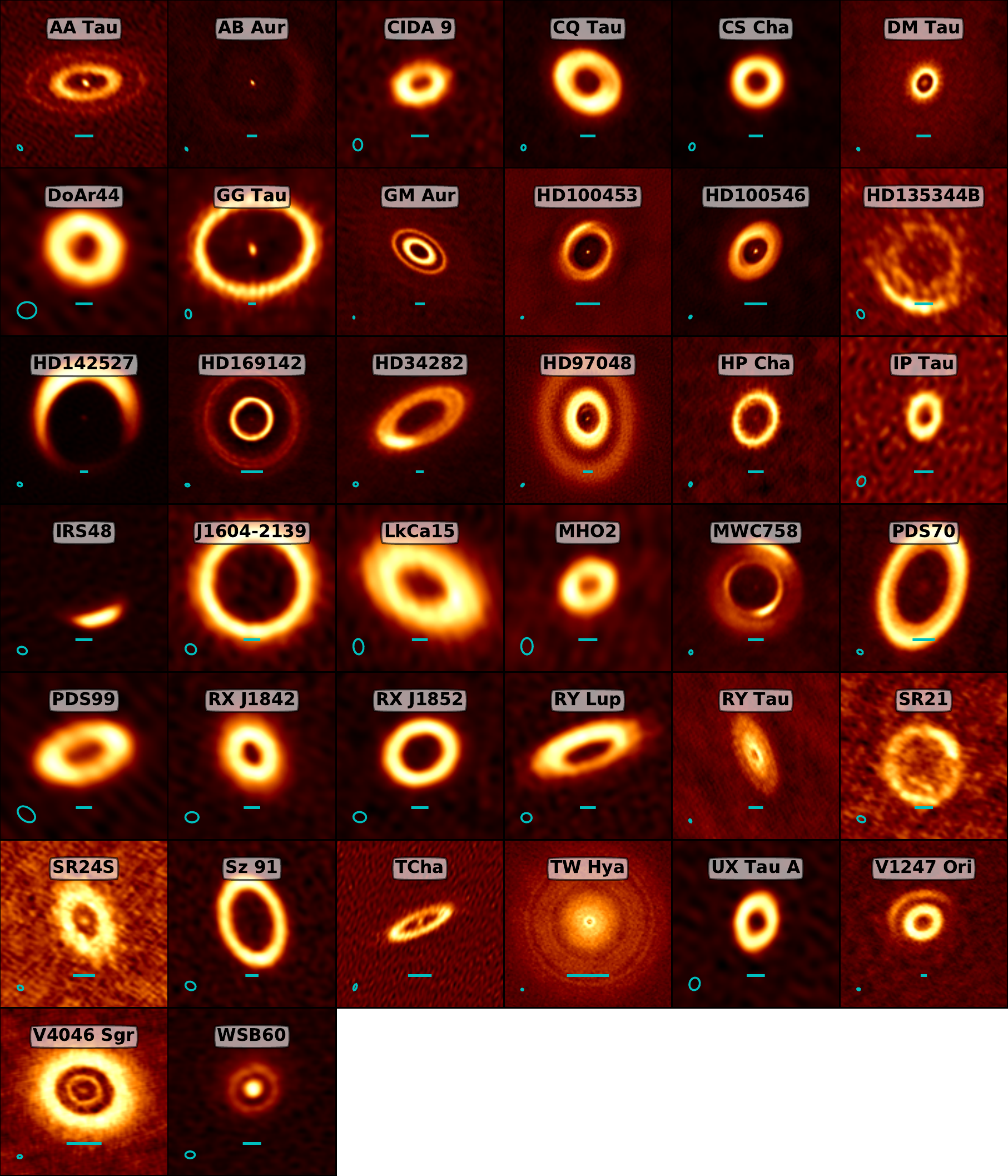}
    \caption{ALMA image continuum gallery of the 38 transition disks in our sample. The size is scaled to the size of the outer disk. The beam size is shown in the bottom left; the scalebar at the bottom is 30 au in length. Note that the outer disk of AB Aur is resolved out due to a lack of short baseline data.}
    \label{fig:image_gallery}
\end{figure*}

\begin{figure*}[htb]
    \centering
    \includegraphics[scale=1.0]{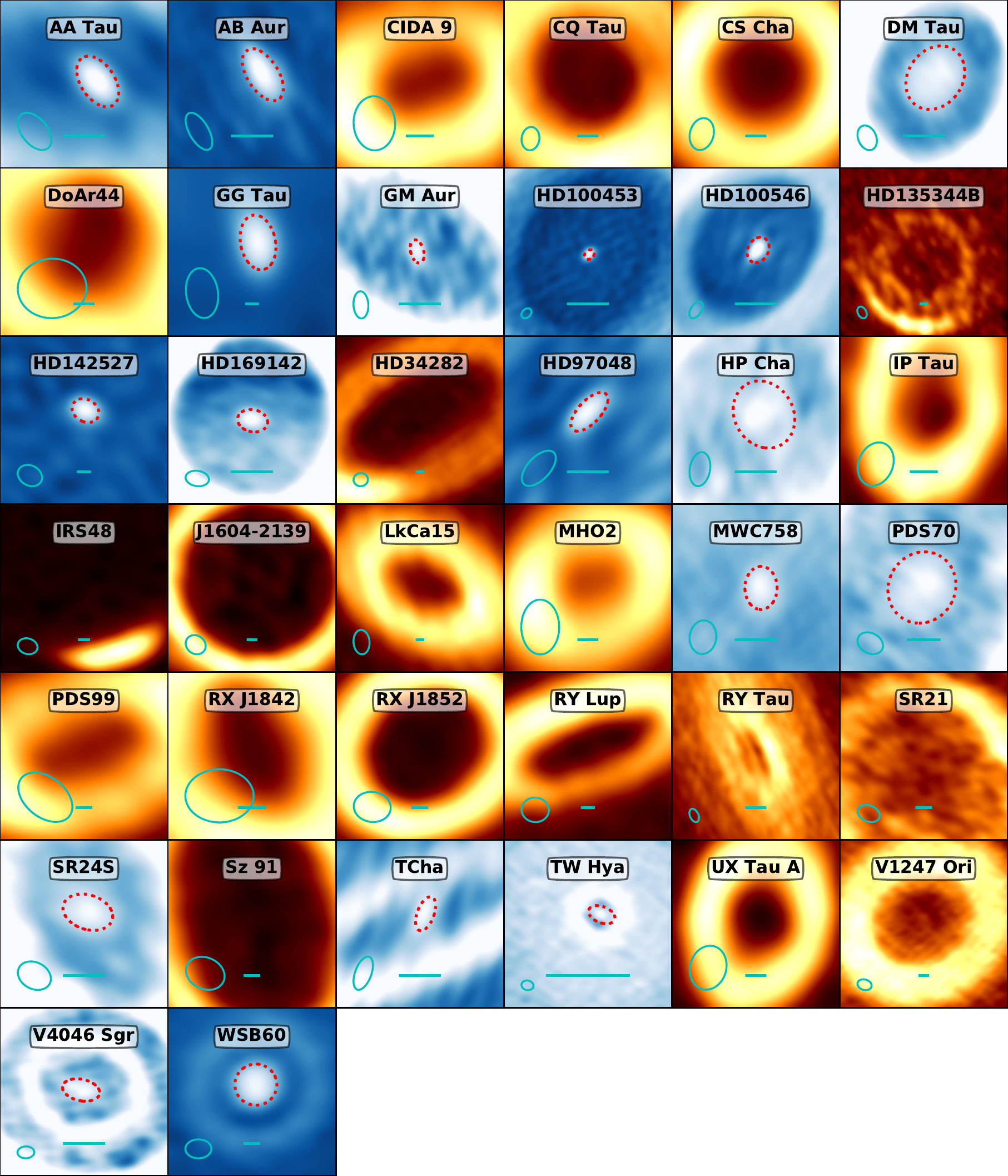}
    \caption{The same gallery as Figure \ref{fig:image_gallery}, but zoomed in on the inner disk. In images where we have detected an inner disk, a blue colormap with the brightest region set to the peak flux of the inner disk is used, and the red dotted line indicates the FWHM of the Gaussian fit. The beam size is shown in the bottom left and the scalebar at the bottom is 10 au in length.} 
    \label{fig:image_gallery_zoom}
\end{figure*}

Images of each target were created using the \texttt{tclean} task in version 5.5.0 of the Common Astronomy Software Applications (CASA) package. For each disk, images were produced using briggs robust weighting values of 2.0 (natural), 0.5 (robust), and -2.0 (uniform). As the relative sensitivity and spatial sampling varied widely across our sample, we individually selected the weighting for each image that maximized  the S/N of the inner disk. For TW Hya, a Briggs robust value of 0.0 was used by \citet{Andrews2016}. The images used for analysis of the inner disk are shown in Figure \ref{fig:image_gallery}, and a zoom-in on each disk with contours in units of the image RMS are shown in Figure \ref{fig:image_gallery_zoom}.

The natural weighted images were used to determine the dust surface density profiles of the outer disk in Section \ref{ssec:outer_disk}. For AB Aur lower resolution Band 7 data (program 2012.1.00303.S) were used to measure the surface density profile of the outer disk, as the Band 6 data did not contain the short baselines and thus did not recover most of the outer disk emission. 

\subsection{Stellar properties}
\label{ssec:stellar_props}

For all targets in our sample spectral types and accretion rates are taken from the literature. For more than half of these targets, these properties were derived using full X-shooter spectra from UV to NIR, with a proper simultaneous fit to UV excess, extinction, stellar photosphere and emission lines following \citet{Alcala2014}. However, most of these stellar studies were performed before Gaia Data Release 2 with incorrect distances for individual objects. We have taken the new distances from Gaia DR2 \citep{Gaia2018} and scaled the luminosities and extinction $A_V$ accordingly by fitting the optical and NIR photometry of the Spectral Energy Distributions to Kurucz stellar photospheric models \citep{CastelliKurucz2004}. Photometric data was taken from \emph{BVRI} surveys, \emph{2MASS} (JHK), \emph{WISE} and \emph{Spitzer-IRAC}. The results are given in Table \ref{tab:stellar} and the final SEDs in Figure \ref{fig:sed_gallery}. For MHO~2 and CIDA~9 (both M stars), no optical photometry is available and the fit to stellar luminosity and extinction remains highly uncertain.

For consistency, we rederive the stellar masses using the new luminosities and temperatures by comparing them with evolutionary tracks for pre-main sequence stars \citep[see method description in ][]{vanderMarel2019}. For K and M stars we use the evolutionary tracks of \citet{Baraffe2015}. For the earlier type stars, the Baraffe models do not provide stellar mass estimates and we use the evolutionary tracks of \citet{Siess2000} instead. 

For the accretion rates, we use the values provided in the literature. These values were derived with pre-Gaia distances but can not be simply scaled without a full consideration of the stellar properties. We assume an uncertainty of half a dex on the accretion rates.

\begin{deluxetable*}{ccccccccccc}
\tablecaption{Stellar Properties}
\label{tab:stellar}
\tablehead{\colhead{Name} & \colhead{Distance} & \colhead{Spectral Type} & \colhead{$T_\text{eff}$} & \colhead{Luminosity} & \colhead{Mass} & \colhead{A$_\textrm{V}$} & \colhead{NIR Excess} & \colhead{PTD/TD} & \colhead{$\dot{M}$} & \colhead{Ref.\tablenotemark{a}}\\ \colhead{ } & \colhead{(pc)} & \colhead{ } & \colhead{(K)} & \colhead{($L_\sun$)} & \colhead{$M_\odot$} & \colhead{ } & \colhead{(\%)} & \colhead{ } & \colhead{($\log(M_\sun/\text{yr}))$} & \colhead{ }}
\startdata
AATau & 137 & K7 & 4350 & 1.1 & 0.68 & 2.3 & $3.3 \pm 3.4$ & TD & -8.44 & 1, 4, 27 \\
ABAur & 163 & A0 & 9520 & 65.1 & 2.56 & 0.5 & $18.9 \pm 1.6$ & PTD & -6.80 & 1, 5, 28 \\
CIDA9 & 140 & M2 & 3580 & 0.1 & 0.36 & 7.0 & $0.2 \pm 3.2$ & TD & 0.00 & 2, 6, - \\
CQTau & 163 & F2 & 6890 & 10.0 & 1.63 & 2.0 & $16.7 \pm 2.1$ & PTD & $<$-8.30 & 1, 7, 29 \\
CSCha & 176 & K2 & 4780 & 1.9 & 1.4 & 1.1 & $0.8 \pm 2.8$ & TD & -8.30 & 1, 8, 8 \\
DMTau & 145 & M2 & 3580 & 0.2 & 0.39 & 0.0 & $1.6 \pm 3.8$ & TD & -8.30 & 1, 8, 30 \\
DoAr44 & 146 & K2 & 4780 & 1.9 & 1.4 & 3.0 & $10.2 \pm 3.2$ & PTD & -8.20 & 1, 8, 8 \\
GGTau AA/Ab & 140 & K7+M0 & 4060 & 1.6 & 0.66 & 0.7 & $13.4 \pm 3.7$ & PTD & -7.30 & 2, 9, 31 \\
GMAur & 160 & K5 & 4350 & 1.0 & 1.01 & 0.3 & $1.8 \pm 3.4$ & TD & -8.30 & 1, 8, 8 \\
HD100453 & 104 & F0 & 7200 & 6.2 & 1.47 & 0.0 & $16.7 \pm 2.0$ & PTD & $<$-8.30 & 1, 10, 10 \\
HD100546 & 110 & A0 & 9520 & 25.0 & 2.13 & 0.0 & $5.7 \pm 0.8$ & PTD & -7.04 & 1, 10, 10 \\
HD135344B & 136 & F5 & 6440 & 6.7 & 1.51 & 0.4 & $21.2 \pm 2.6$ & PTD & -7.37 & 1, 10, 10 \\
HD142527 & 157 & F6 & 6360 & 9.9 & 1.69 & 0.3 & $51.1 \pm 4.7$ & PTD & -7.45 & 1, 10, 10 \\
HD169142 & 114 & A5 & 8200 & 8.0 & 1.65 & 0.4 & $6.6 \pm 1.2$ & PTD & -8.70 & 1, 11, 32 \\
HD34282 & 312 & A0 & 9520 & 10.8 & 2.11 & 0.2 & $16.9 \pm 1.5$ & PTD & $<$-8.30 & 1, 10, 10 \\
HD97048 & 185 & A0 & 9520 & 30.0 & 2.17 & 0.9 & $11.8 \pm 1.2$ & PTD & $<$-8.16 & 1, 10, 10 \\
HPCha & 160 & K7 & 4060 & 2.4 & 0.95 & 1.5 & $43.3 \pm 4.7$ & PTD & -8.97 & 3, 12, 12 \\
IPTau & 131 & M0 & 3850 & 0.6 & 0.54 & 1.7 & $4.0 \pm 3.9$ & PTD & -8.14 & 1, 13, 33 \\
IRS48 & 134 & A0 & 9520 & 17.8 & 1.96 & 11.0 & $3.8 \pm 0.7$ & PTD & -8.40 & 1, 14, 34 \\
J1604.3-2130 & 150 & K3 & 4780 & 0.7 & 1.1 & 0.5 & $0.9 \pm 2.0$ & PTD & -10.54 & 1, 15, 35 \\
LkCa15 & 159 & K2 & 4730 & 1.3 & 1.32 & 1.2 & $4.5 \pm 2.9$ & PTD & -8.40 & 1, 8, 8 \\
MHO2 & 133 & M3 & 3470 & 1.0 & 0.44 & 0.0 & $0.0 \pm 0.7$ & TD & 0.00 & 1, 16, - \\
MWC 758 & 160 & A7 & 7850 & 14.0 & 1.77 & 0.7 & $14.8 \pm 1.6$ & PTD & -7.35 & 1, 17, 34 \\
PDS70 & 113 & K7 & 4060 & 0.3 & 0.8 & 0.0 & $4.2 \pm 3.8$ & TD & $<$-11.00 & 1, 18, 36 \\
PDS99 & 155 & K6 & 4205 & 1.1 & 0.88 & 2.0 & $0.9 \pm 3.4$ & TD & 0.00 & 1, 19, - \\
RXJ1842.9-3532 & 154 & K2 & 4780 & 0.8 & 1.14 & 0.4 & $17.3 \pm 3.7$ & PTD & -8.80 & 1, 8, 8 \\
RXJ1852.3-3700 & 146 & K2 & 4780 & 0.6 & 1.05 & 0.7 & $0.8 \pm 2.8$ & TD & -8.70 & 1, 8, 8 \\
RYLup & 159 & K2 & 4780 & 1.9 & 1.4 & 0.4 & $32.9 \pm 4.6$ & PTD & -8.20 & 1, 20, 20 \\
RYTau & 175 & G2 & 5860 & 15.0 & 2.25 & 2.2 & $35.8 \pm 4.0$ & PTD & -7.10 & 1, 21, 21 \\
SR21 & 138 & G4 & 5770 & 11.0 & 2.12 & 6.0 & $0.4 \pm 1.7$ & PTD & -7.90 & 1, 8, 8 \\
SR24S & 114 & K6 & 4060 & 2.5 & 0.87 & 8.0 & $11.7 \pm 3.7$ & PTD & -7.15 & 1, 22, 22 \\
Sz91 & 159 & M1 & 3850 & 0.2 & 0.54 & 1.2 & $1.6 \pm 3.8$ & TD & -8.73 & 1, 20, 14 \\
Tcha & 107 & G8 & 5570 & 1.3 & 1.12 & 2.0 & $27.8 \pm 3.4$ & PTD & -8.40 & 1, 23, 37 \\
TWHya & 60 & K7 & 4205 & 0.3 & 0.81 & 0.0 & $4.1 \pm 3.6$ & TD & -8.90 & 1, 8, 8 \\
UXTauA & 140 & G8 & 5570 & 2.5 & 1.4 & 1.4 & $6.8 \pm 2.4$ & PTD & -7.95 & 1, 24, 24 \\
V1247Ori & 400 & F0 & 7200 & 15.0 & 1.82 & 0.0 & $21.1 \pm 2.2$ & PTD & -8.00 & 1, 25, 38 \\
V4046Sgr & 72 & K7+K5 & 4060 & 0.5 & 0.76 & 0.0 & $0.9 \pm 3.7$ & TD & -9.30 & 1, 26, 39 \\
WSB60 & 137 & M6 & 3050 & 0.2 & 0.24 & 3.7 & $36.2 \pm 4.8$ & PTD & -8.90 & 1, 8, 8
\enddata
\tablecomments{{a) References order: distance, spectral type, $\dot{M}$.}
{\bf Refs.}
% distance refs
1) \citet{Gaia2018}; % All except CIDA9, GG Tau, HP Cha
2) \citet{Kenyon2008}; % 140pc to CIDA 9 and GG Tau in Taurus
3) \citet{Feigelson2004}; % average 160pc distance to Chameleon, used for HP Cha
% Spectral type refs
4) \citet{Bouvier1999}; % AA Tau
5) \citet{Bohm1993}; % AB Aur
6) \citet{Herczeg2014}; % CIDA 9
7) \citet{Herbig1960}; % CQ Tau
8) \citet{Manara2014}; % CS Cha
%8) \citet{Manara2014}; % DM Tau
%8) \citet{Manara2014}; % DoAr44
9) \citet{White1999}; % GG Tau 
%8) \citet{Manara2014}; % GM Aur
10) \citet{Fairlamb2015}; % HD100453
%10) \citet{Fairlamb2015}; % HD100546
%10) \citet{Fairlamb2015}; % HD135344B
%10) \citet{Fairlamb2015}; % HD142527
11) \citet{Dunkin1997}; % HD169142
%10) \citet{Fairlamb2015}; % HD34282
%10) \citet{Fairlamb2015}; % HD97048
12) \citet{Manara2017}; % HP Cha
13) \citet{Furlan2011}; % IP Tau
14) \citet{Brown2012a}; % IRS48
15) \citet{Sicilia-Aguilar2020} % J1604
%8) \citet{Manara2014}; % LkCa15
16) \citet{Luhman2010}; % MHO2
17) \citet{Acke2005}; % MWC758
18) \citet{Pecaut2016}; % PDS70
19) \citet{Torres2006}; % PDS99
%8) \citet{Manara2014}; % RXJ1842
%8) \citet{Manara2014}; % RXJ1852
20) \citet{Alcala2017}; % RY Lup
21) \citet{Calvet2004}; % RY Tau
%8) \citet{Manara2014}; % SR21
22) \citet{Natta2006}; % SR24S
%20) \citet{Alcala2017}; % Sz91
23) \citet{Alcala1993}; % TCha
%8) \citet{Manara2014}; % TW Hya
24) \citet{Espaillat2010}; % UX Tau A
25) \citet{Kraus2013}; % V1247 Ori
26) \citet{Stempels2004}; % V4046Sgr
%8) \citet{Manara2014}; % WSB60
% accretion rate refs
27) \citet{Bouvier2013};	%	AATau
28) \citet{GarciaLopez2006};	%	ABAur
 %No accretion rate!		CIDA9
29) \citet{Mendigutia2011};	%	CQTau
%8) \citet{Manara2014};	%	CSCha
30) \citet{Rigliaco2015}; 	%	DMTau
%8) % Manara et al. 2014	%	DoAr44
31) \citet{Beck2012};	%	GGTau AA/Ab
%8) % Manara et al. 2014	%	GMAur
%10) \citet{Fairlamb2015};	%	HD100453
%10) %Fairlamb et al. 2015	%	HD100546
%10) %Fairlamb et al. 2015	%	HD135344B
%10) %Fairlamb et al. 2015	%	HD142527
32) \citet{Wagner2015};	%	HD169142
%10) %Fairlamb et al. 2015	%	HD34282
%10) %Fairlamb et al. 2015	%	HD97048
%12) \citet{Manara2017}; 	%	HPCha
33) \citet{Ingleby2013};	%	IPTau
34) \citet{Salyk2013};	%	IRS48
35) \citet{Pinilla2018}; % (from an in prep manara paper)	%	J1604.3-2130
%8) % Manara et al. 2014	%	LkCa15
%- %No accretion rate!	%	MHO2
%34) %Salyk et al. 2013	%	MWC 758
36) \citep{Wagner2018}	%	PDS70
%-	%	No accretion rate!PDS99
%8) %Manara et al. 2014	%	RXJ1842.9-3532
%8) %Manara et al. 2014	%	RXJ1852.3-3700
%20) \citet{Alcala2017};	%	RYLup
%21) \citet{Calvet2004};	%	RYTau
%8) %Manara et al. 2014	%	SR21
%22) \citet{Natta2006}; 	%	SR24S
%14) %Alcala et al. 2017	%	Sz91
37) \citet{Schisano2009};	%	Tcha
%8) %Manara et al. 2014	%	TWHya
%24) \citet{Espaillat2010};	%	UXTauA
38) \citet{Willson2019};	%	V1247Ori
39) \citet{Curran2011}.	%	V4046Sgr
%8) %Manara et al. 2014	%	WSB60
}
\end{deluxetable*}

\subsection{NIR excess}
\label{ssec:nir_excess_calc}
To quantify the presence of warm micron-sized dust close to the star, we compute the percentage near-infrared (NIR) following the definition from \citet{Pascual2016}.
Specifically, we calculate the percentage NIR excess from the de-reddened SED excess as $P_\text{NIR}=100\times F_\text{NIR}/F_\star$, where $F_\text{NIR}$ and $F_\star$ are the flux of the SED and stellar photosphere integrated between 1.2 to 4.5 \micron, respectively. As the uncertainty in $P_\text{NIR}$ is dominated by variations in $A_\text{V}$, we determine it by assuming a 1 $\sigma$ uncertainty of 0.5 in $A_\text{V}$ and propagating errors accordingly. In addition, we classify each disk as a pre-transition disk (PTD), or otherwise a transition disk (TD) using the K-band (2.2 \micron) excess of the SED over the stellar photosphere \citep{Espaillat2007}. If $\Delta K > 0.25$ magnitudes, the disk is classified as pre-transition, and transition otherwise. The percentage NIR excess and PTD/TD status of each disk are listed in Table \ref{tab:stellar}. Although the comparison is not entirely the same throughout the sample, the threshold for a transition to a pre-transition disk appears to be around a NIR excess of $\sim3\%$.

%\clearpage\newpage

\section{Analysis}
\label{sec:analysis}

\subsection{Inner disk detection and modelling}
\label{ssec:inner_fits}
We detect an inner disk in 18 of the 38 images. A detection is defined as the measurement of a flux at the stellar location above a threshold of 3 times the RMS image noise level. For each of the 18 detections, a Gaussian model is fit using the CASA \texttt{imfit} task. Otherwise, we use 3 times the RMS as the upper limit on the flux for an unresolved inner disk. The results of our inner disk fits and upper limits are given in Table \ref{tab:inner_outer_disks}, while the RA and Dec. of the inner disks can be found in Table \ref{tab:inner_disk_coords} in the Appendix.

\begin{figure}[htb]
    \centering
    \includegraphics[scale=1.0]{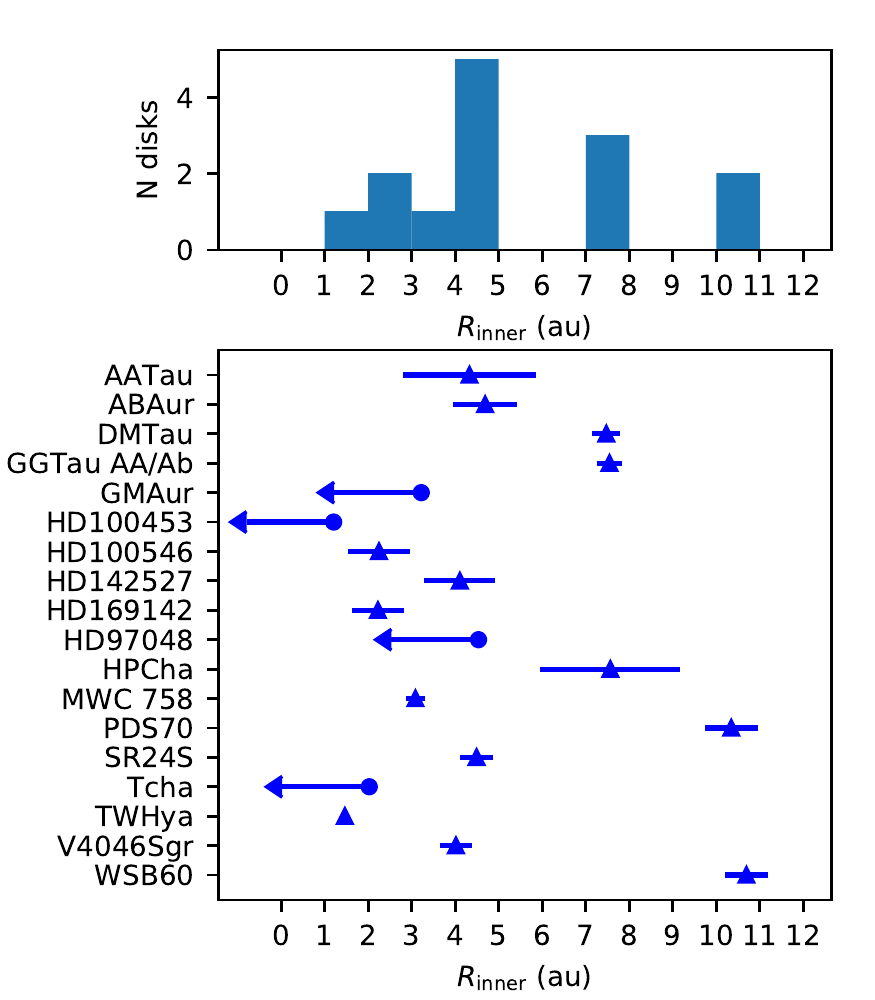}
    \caption{Histogram of the outer radius of the resolved inner disks. Resolved measurements are shown with a blue triangle, unresolved disks with a blue circle. Upper limit contributions to the histogram are shown with a red cross-hatch.} 
    \label{fig:rgap_hist}
\end{figure}

Of the 18 detected inner disks, we are able to resolve 14 of them, and measure the inclination and outer radius of the inner disk ($R_\text{inner}$). To determine $R_\text{inner}$ we measure the half width at half maximum (HWHM) of the Gaussian fit in the resolved case; where unresolved, we assume the inner disk has the same inclination and position angle as the outer disk, and use the semi-major axis of the largest ellipse (i.e., the largest projected inner disk) that would fit into the beam as $R_\text{inner}$. The inclinations and position angle of the inner disk are given in Table \ref{tab:misalignment}, while the values of $R_\text{inner}$ scaled to au are listed in Table \ref{tab:inner_outer_disks}. In Figure \ref{fig:rgap_hist}, we show the distribution of $R_\text{inner}$ in our sample, which has a mean value of $\sim$5~au.   

\begin{deluxetable*}{ccccccccc}
\label{tab:inner_outer_disks}
\tablecaption{Inner and Outer Disk Properties}
\tablehead{\colhead{Name} & \colhead{Inner Flux} & \colhead{Inner Dust Mass} & \colhead{Maj. x Min.} & \colhead{$R_\text{inner}$} & \colhead{$R_\text{cav}$} & \colhead{$R_\text{outer}$} & \colhead{Outer Flux} & \colhead{$\delta_\text{dust}$}\\ \colhead{ } & \colhead{(mJy)} & \colhead{ $M_\earth$} & \colhead{(\arcsec)} & \colhead{ } & \colhead{ } & \colhead{ } & \colhead{(mJy)} & \colhead{ }}
\startdata
AATau & $1.91 \pm 0.30$ & $0.071 \pm 0.036$ & 0.06 x 0.04 & $4.3 \pm 1.5$ & 44 & 111 & 67 & -1.43 \\
ABAur & $2.16 \pm 0.17$ & $0.057 \pm 0.012$ & 0.06 x 0.03 & $4.7 \pm 0.7$ & 156 & 225 & 69 & -1.47 \\
CIDA9 & $<0.21$ & $<0.032$ & - & - & 29 & 46 & 35 & -2.41 \\
CQTau & $<0.06$ & $<0.005$ & - & - & 50 & 70 & 147 & -3.48 \\
CSCha & $<0.12$ & $<0.004$ & - & - & 37 & 72 & 186 & -3.09 \\
DMTau & $1.50 \pm 0.06$ & $0.185 \pm 0.014$ & 0.10 x 0.08 & $7.5 \pm 0.3$ & 25 & 137 & 119 & -1.59 \\
DoAr44 & $<0.99$ & $<0.025$ & - & - & 40 & 88 & 192 & -2.09 \\
GGTau AA/Ab & $27.34 \pm 0.20$ & $0.429 \pm 0.031$ & 0.11 x 0.06 & $7.5 \pm 0.3$ & 224 & 258 & 2047 & -2.49 \\
GMAur & $0.05 \pm 0.01$ & $<0.004$ & $<$ 0.04 x 0.02 & $< 3.2$ & 40 & 220 & 162 & -2.73 \\
HD100453 & $1.29 \pm 0.11$ & $<0.013$ & $<$ 0.03 x 0.02 & $< 1.2$ & 30 & 31 & 200 & -3.09 \\
HD100546 & $5.20 \pm 0.90$ & $0.097 \pm 0.033$ & 0.04 x 0.04 & $2.2 \pm 0.7$ & 27 & 99 & 432 & -1.24 \\
HD135344B & $<0.08$ & $<0.037$ & - & - & 52 & 86 & 200 & -3.36 \\
HD142527 & $3.70 \pm 0.16$ & $0.061 \pm 0.018$ & 0.05 x 0.04 & $4.1 \pm 0.8$ & 185 & 245 & 967 & -3.53 \\
HD169142 & $0.31 \pm 0.04$ & $0.007 \pm 0.002$ & 0.04 x 0.03 & $2.2 \pm 0.6$ & 26 & 127 & 198 & -2.12 \\
HD34282 & $<0.06$ & $<0.017$ & - & - & 87 & 227 & 100 & -2.67 \\
HD97048 & $2.84 \pm 0.34$ & $<0.062$ & $<$ 0.06 x 0.03 & $< 4.5$ & 63 & 189 & 2344 & -2.33 \\
HPCha & $0.61 \pm 0.09$ & $0.056 \pm 0.016$ & 0.09 x 0.08 & $7.6 \pm 1.6$ & 50 & 64 & 66 & -2.68 \\
IPTau & $<0.22$ & $<0.019$ & - & - & 25 & 31 & 13 & -2.27 \\
IRS48 & $<0.71$ & $<0.009$ & - & - & 83 & 98 & 173 & -2.71 \\
J1604.3-2130 & $<0.43$ & $<0.013$ & - & - & 87 & 135 & 262 & -2.72 \\
LkCa15 & $<0.47$ & $<0.025$ & - & - & 76 & 114 & 252 & -2.73 \\
MHO2 & $<1.19$ & $<0.097$ & - & - & 28 & 55 & 161 & -2.08 \\
MWC 758 & $0.34 \pm 0.01$ & $0.003 \pm 0.000$ & 0.04 x 0.02 & $3.1 \pm 0.2$ & 62 & 102 & 217 & -2.75 \\
PDS70 & $2.02 \pm 0.11$ & $0.047 \pm 0.005$ & 0.18 x 0.16 & $10.3 \pm 0.6$ & 74 & 87 & 287 & -2.82 \\
PDS99 & $<0.41$ & $<0.043$ & - & - & 56 & 81 & 89 & -2.44 \\
RXJ1842.9-3532 & $<0.37$ & $<0.012$ & - & - & 37 & 73 & 133 & -2.45 \\
RXJ1852.3-3700 & $<0.51$ & $<0.017$ & - & - & 49 & 74 & 149 & -2.57 \\
RYLup & $<0.17$ & $<0.018$ & - & - & 69 & 80 & 65 & -3.07 \\
RYTau & $<0.20$ & $<0.016$ & - & - & 27 & 57 & 271 & -3.00 \\
SR21 & $<0.05$ & $<0.024$ & - & - & 56 & 66 & 7 & -2.57 \\
SR24S & $0.48 \pm 0.03$ & $0.119 \pm 0.018$ & 0.08 x 0.04 & $4.5 \pm 0.4$ & 35 & 42 & 25 & -2.44 \\
Sz91 & $<0.13$ & $<0.005$ & - & - & 86 & 101 & 46 & -2.97 \\
Tcha & $0.29 \pm 0.03$ & $<0.119$ & $<$ 0.08 x 0.08 & $< 2.0$ & 34 & 46 & 17 & -1.82 \\
TWHya & $2.29 \pm 0.03$ & $0.004 \pm 0.000$ & 0.03 x 0.02 & $1.0 \pm 0.0$ & 2 & 57 & 1319 & -2.13 \\
UXTauA & $<0.98$ & $<0.020$ & - & - & 31 & 48 & 161 & -2.36 \\
V1247Ori & $<0.18$ & $<0.020$ & - & - & 64 & 126 & 259 & -2.87 \\
V4046Sgr & $1.17 \pm 0.09$ & $0.013 \pm 0.002$ & 0.11 x 0.06 & $4.0 \pm 0.4$ & 31 & 45 & 248 & -2.83 \\
WSB60 & $23.65 \pm 0.67$ & $3.559 \pm 0.269$ & 0.16 x 0.14 & $10.7 \pm 0.5$ & 32 & 72 & 90 & -0.54
\enddata
\end{deluxetable*}

%\newpage
\subsection{Inner disk dust mass}
\label{ssec:inner_dust_mass}

In order to derive the inner disk mass from the millimeter flux, we use the following approach. First, we check the optical depth by estimating the expected flux based on basic radiative transfer calculations, given the stellar luminosity.

In general, the flux density $F_{\nu}$ is defined
\begin{equation}
\label{eqn:flux}
F_{\nu} = \int I_{\nu}d\Omega = \frac{1}{d^2}\int_{R_\text{in}}^{R_\text{out}} I_{\nu}2\pi rdr
\end{equation}
for distance $d$ and assuming the emission originates from an annulus with an inner radius $R_\text{in}$ and outer radius $R_\text{out}$.

The specific intensity $I_{\nu}$ is defined as
\begin{equation}
\label{eqn:intensity}
I_{\nu} = B_{\nu}(T(r))(1-e^{-\tau(r)})
\end{equation}
with $\tau(r)$ the optical depth and $B_{\nu}(T(r))$ the Planck equation in the Rayleigh-Jeans approximation:
\begin{equation}
B_{\nu}(T(r)) = \frac{2\nu^2 k_BT(r)}{c^2}
\end{equation}
and the midplane temperature profile with the simplified expression for a passively heated, flared disk in radiative equilibrium \citep[e.g.][]{ChiangGoldreich1997,Dullemond2001}:
\begin{equation}
\label{eqn:temperature}
T(r) = \Big(\frac{\phi L_*}{8\pi\sigma_Br^2}\Big)^{1/4} = \sqrt[4]{\frac{\phi L_*}{8\pi\sigma_B}} \frac{1}{\sqrt{r}}
\end{equation}
with $\sigma_B$ the Stefan-Boltzmann constant, $\phi$ the flaring angle (taken as 0.02) and $L_*$ the stellar luminosity.

For an entirely optically thick disk, $\tau(r)\gg1$, so $1-e^{-\tau(r)}\approx1$, and Equation \ref{eqn:flux} simplifies to:\\
\begin{align}
\label{eqn:flux_thick}
F_{\nu} 
&= \frac{1}{d^2}\frac{2\nu^2 k_B}{c^2} \sqrt[4]{\frac{\phi L_*}{8\pi\sigma_B}} 2\pi \int_{R_\text{in}}^{R_\text{out}} \frac{1}{\sqrt{r}} rdr \nonumber \\
&= \frac{1}{d^2}\frac{2\nu^2 k_B}{c^2} \sqrt[4]{\frac{\phi L_*}{8\pi\sigma_B}} \frac{4}{3}\pi (R_\text{out}^{3/2}-R_\text{in}^{3/2})
\end{align}

However, if the emission is optically thin, $1-e^{-\tau(r)}\approx \tau(r)$, and  $\tau(r)$ in Equation \ref{eqn:intensity} is defined for a geometrically thin disk as:
\begin{equation}
\tau(r) = \frac{\kappa_{\nu}\Sigma_d(r)}{\cos i} 
\end{equation}
with the dust opacity $\kappa_{\nu}$. We thus assume a power law for the dust surface density, 

\begin{equation}
\label{eqn:dust_pwr}
\Sigma_d(r) = \Sigma_c\big(\frac{r}{r_c}\big)^{-1} = \Sigma_c\frac{r_c}{r}
\end{equation}

and Eqn \ref{eqn:flux} then becomes

\begin{align}
\label{eqn:flux_thin}
F_{\nu} 
&= \frac{1}{d^2}\frac{2\nu^2 k_B}{c^2}\sqrt[4]{\frac{\phi L_*}{8\pi\sigma_B}} 2\pi \frac{\kappa_{\nu}\Sigma_c}{\cos i}r_c \nonumber \int_{R_\text{in}}^{R_\text{out}}  \frac{1}{\sqrt{r}} r^{-1} rdr \\
&= \frac{1}{d^2}\frac{2\nu^2 k_B}{c^2} \sqrt[4]{\frac{\phi L_*}{8\pi\sigma_B}} 4\pi \frac{\kappa_{\nu}\Sigma_c}{\cos i}r_c (R_\text{out}^{1/2}-R_\text{in}^{1/2}).
\end{align}

We then choose the scaling surface density and radius to be evaluated at the outer radius, i.e., $r_c=R_\text{out}$ and $\Sigma_c=\Sigma_\text{out}$, such that

\begin{equation}
F_\nu = \frac{1}{d^2}\frac{2\nu^2 k_B}{c^2} \sqrt[4]{\frac{\phi L_*}{8\pi\sigma_B}} 4\pi \tau_\text{out}R_\text{out}(R_\text{out}^{1/2}-R_\text{in}^{1/2}), 
\end{equation}

where $\tau_\text{out}$ is the optical depth at the outer radius.

\begin{figure*}[htb]
    \centering
    \includegraphics[scale=1.0]{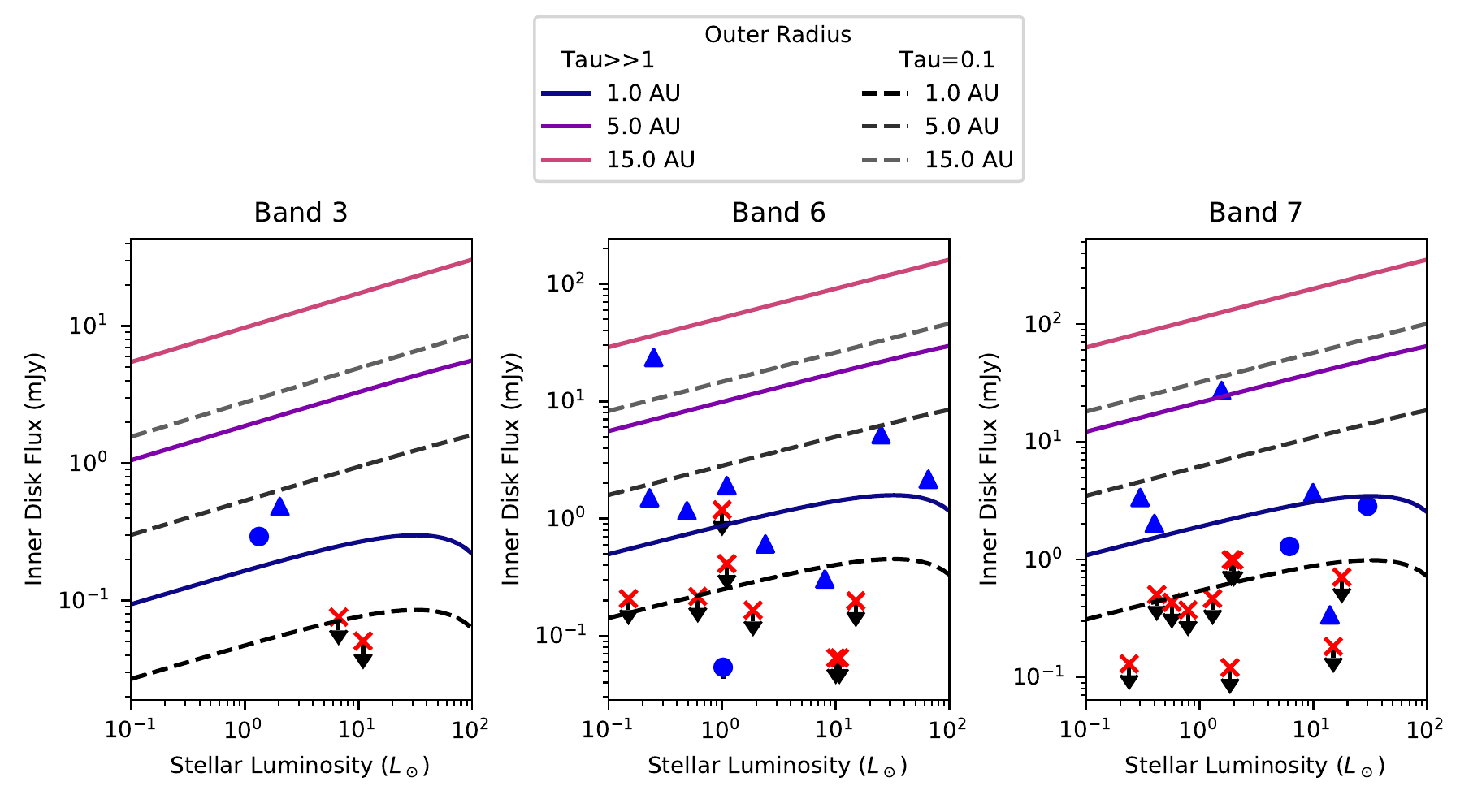}
    \caption{Predicted inner disk flux as a function of $L_*$ for a variety of $R_\text{inner}$ in the optically thick and optically thin cases, and for three different frequencies. The data from our sample are overplotted: the detections in blue (triangles are resolved, circles unresolved) and the upper limits in red. All but two resolved fluxes are below the optically thick limit for a size of 5 au, and most unresolved fluxes are below this limit for a size of 1 au.} 
    \label{fig:inner_disk_opacity}
\end{figure*}

The flux thus depends primarily on $\tau_\text{out}$ and $L_*$. For the inner disk, we define the outer radius $R_\text{out}=R_\text{inner}$, and the inner radius as the sublimation radius $R_\text{in}=R_\text{sub}=0.07\sqrt{L_*(L_{\odot})}$ \citep{Dullemond2001}. 
 
Figure \ref{fig:inner_disk_opacity} shows the expected flux as function of $L_*$ for both the optically thick and optically thin ($\tau_\text{out}=0.1$) case for a range of outer radii in the ALMA bands of our observations. The measured inner disk fluxes are overplotted. In order to assess optical depth, the outer radius ($R_\text{inner}$) thus needs to be known.

For our data $R_\text{inner}$ is constrained for 14 targets. For the unresolved inner disks and the non-detections the outer radius is not constrained, but based on the resolved cases (Figure \ref{fig:rgap_hist}), we assume an average outer radius of 5 au for those inner disks. However, Figure \ref{fig:inner_disk_opacity} shows that the emission remains optically thin for all but 2 disks if $R_\text{inner}$=5 au. These disks are WSB 60 and GG Tau, which are resolved with $R_\text{inner}=10.7\pm0.5$ and $R_\text{inner}=7.5\pm0.5$ au, and are thus within the optically thin regime as well. 

This means that the mm-dust mass can computed accurately with the optically thin assumption. We note that in this approach using the blackbody approximation we ignore scattering, which might result in an underestimate of the inner disk dust mass \citep{Zhu2019}. However, a full radiative transfer modeling approach is beyond the scope of this study.

The dust mass is defined as
\begin{equation}
M_\text{dust} = \int_{R_\text{in}}^{R_\text{out}} \Sigma_d(r)2\pi rdr
\end{equation}

resulting in

\begin{equation}
\label{eqn:sigma}
M_\text{dust} 
= 2\pi\Sigma_cr_c ({R_\text{out}}-{R_\text{in}})
\end{equation}

using the power law profile in Eqn \ref{eqn:dust_pwr}. Now we can express the optically thin flux from Equation \ref{eqn:flux_thin} in terms of the dust mass:

\begin{align*}
F_{\nu}
&=\frac{1}{d^2}\frac{2\nu^2 k_B}{c^2} \sqrt[4]{\frac{\phi L_*}{8\pi\sigma_B}}  \frac{\kappa_{\nu}}{\cos i} M_\text{dust} \frac{2({R_\text{out}^{1/2}-{R_\text{in}^{1/2}}})}{({R_\text{out}}-{R_\text{in}})}  
\end{align*}
{\rm or}
\begin{equation}
M_\text{dust} = \frac{F_{\nu}d^2 \cos i}{\kappa_{\nu} \frac{2\nu^2 k_B}{c^2} \sqrt[4]{\frac{\phi L_*}{8\pi\sigma_B}}  }\frac{({R_\text{out}}-{R_\text{in}})}{2({R_\text{out}^{1/2}-{R_\text{in}^{1/2}}})},
\label{eqn:final}
\end{equation}

where we take $\kappa_{\nu}$ as 10 cm$^2$ g$^{-1}$ at 1000 GHz and use an opacity power-law index of $\beta$ = 1.0 \citep{Beckwith1991}.

In this calculation we assume that there are no significant contributions from free-free emission from ionized gas close to the star from a jet \citep{Snell1986}. Considering that most detections are resolved, this suggests that for those disks the main component is originating from thermal dust emission from a disk. Although free-free emission at 9 GHz has been detected for a handful of our targets \citep{Zapata2017}, extrapolation of these fluxes to our observed wavelengths remains highly uncertain. Multi-wavelength observations of the inner disk are required to measure the spectral index, which can rule out contributions of free-free emission. If free-free emission is contributing to our derived fluxes, this implies that even our inner disk masses from detections are upper limits.

\subsection{Disk surface density profiles}
\label{ssec:outer_disk}
Using the total flux of the outer and inner disk, we can compute $\Sigma_cr_c$ from Equation \ref{eqn:sigma} and \ref{eqn:final} for both the outer and inner disk and construct $\Sigma_d(r)$ using the parameter $\delta_\text{dust}$, the fractional drop in dust surface density from the inner disk with respect to the outer disk:

\begin{align*}
\Sigma_d(r) &= \frac{\Sigma_c r_c}{r} & \textrm{ for } r>R_\text{cav} \\
&= 0 & \textrm{ for } R_\text{inner}<r<R_\text{cav} \\
&= \delta_\text{dust}\frac{\Sigma_c r_c}{r} & \textrm{ for } R_\text{sub}<r<R_\text{inner} 
    \label{eqn:ddust}
\end{align*}

\begin{figure}[htb]
    \centering
    \includegraphics[]{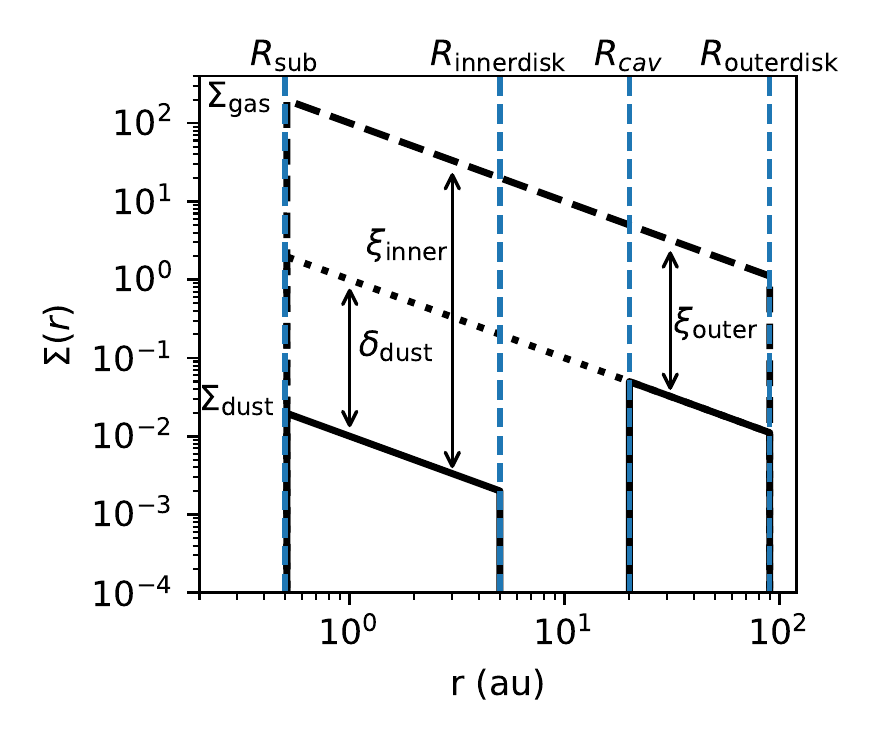}
    \caption{Generic surface density profile of gas and dust. The profile is described by a power law $\Sigma(r) = \Sigma_cr_c/r$, and the drop in the inner dust disk is described by $\delta_\text{dust}$. The inner and outer radius of inner and outer disk are described by $R_\text{sub}, R_\text{inner}, R_\text{cav}$ and $R_\text{outer}$, respectively. The gas surface density profile is described by the gas-to-dust ratios $\xi_\text{inner}$ and $\xi_\text{outer}$.} 
    \label{fig:genericprofile}
\end{figure}

For the outer disk, we take $R_\text{in}$ = $R_\text{cav}$ from the azimuthally averaged intensity profile to be representative of the cavity radius. In reality the cavity radius edge is usually more complex than a sharp edge \citep[see e.g.][]{Pinilla2018} but as we only aim for an approximate profile of $\Sigma_d(r)$ compared to the inner disk, this approach is sufficient. Our derived values for $R_\text{cav}$ are generally within 20\% of the literature values of $R_\text{cav}$ derived using more detailed fitting methods. 

For $R_\text{out}$, we use a curve-of-growth method to measure the radius of the outer disk $R_\text{outer}$, in which successively larger photometric apertures are applied until the measured flux is 95\% of the total flux. A generic profile is shown in Figure \ref{fig:genericprofile}, listing the definitions of the different radii and ratios in the disk. The outer disk properties are given in Table \ref{tab:inner_outer_disks}. We also provide the inclination and position angle of the outer disk in Table \ref{tab:misalignment}, taken from the literature where available (references in Table \ref{tab:disk_sample}). For CS~Cha, HP~Cha, MHO~2, PDS~99, RXJ1842.9-3532, UX~Tau A, and WSB~60 no previous fitting of the outer disk was performed. We estimate the inclination and position angle by fitting a simple Gaussian ring to the outer disk ring (see Appendix). 

\begin{figure*}[htb]
    \centering
    \includegraphics[scale=1.0]{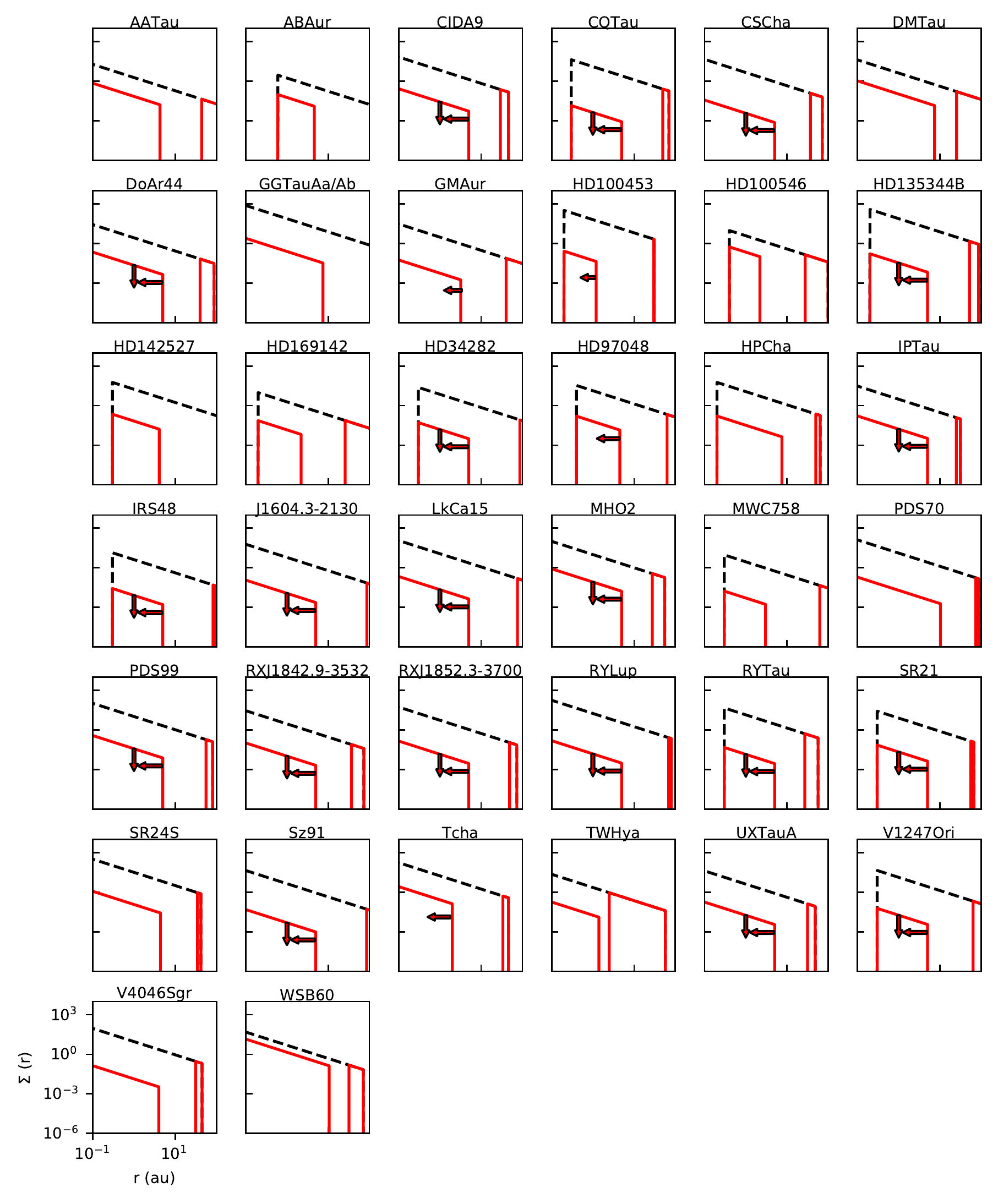}
    \caption{Derived dust surface density profiles for each disk, computing the dust mass in inner and outer disk, using the inner and outer radii and using the relations derived in Section \ref{ssec:outer_disk}. Arrows indicate whether the inner dust disk is a non-detection (vertical) and/or unresolved (horizontal). For the non-detections, an average size $R_\text{inner}$=5 au is assumed. Each inner disk is depleted with respect to the outer disk ($\delta_\text{dust}$).} 
    \label{fig:surfdensity}
\end{figure*}

The final $\Sigma_d(r)$ profiles are given in Figure \ref{fig:surfdensity}. For unresolved and undetected fluxes, we provide the profile for the assumption $R_\text{inner}=5$ au (arrows indicating that these are upper limits). For all inner disks, the surface density drops by 2 to 3 orders of magnitude compared to the unperturbed surface density profile extrapolated from the outer disk, this drop is quantified as $\delta_\text{dust}$ in Table \ref{tab:inner_outer_disks}.

\section{Discussion}
\label{sec:discussion}

\subsection{Misalignment between the inner and outer disk}
\label{ssec:misalignment_disc}

\begin{figure}[htb]
    \centering
    \includegraphics[scale=1.0]{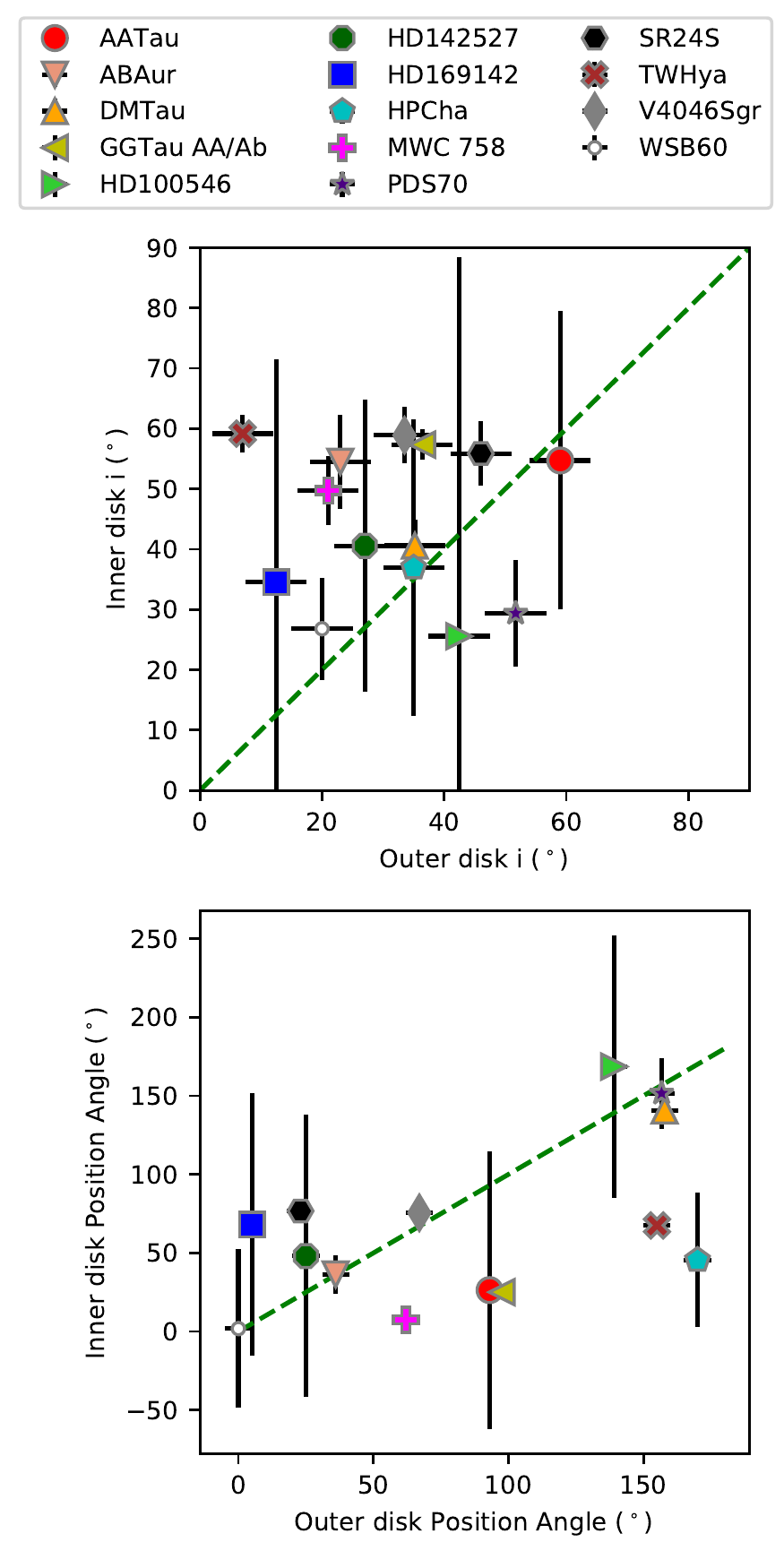}
    \caption{Comparison of the orientation of inner and outer dust disks in inclination and position angle for the resolved inner dust disks in our sample. Each symbol and color represents a different target (legend on top). For most inner disks, either position angle or inclination or both are significantly different from the outer disk, indicating a misalignment.} 
    \label{fig:misalign_i_pa}
\end{figure}

For 14 of 18 disks in our sample, we are able to resolve the inner disks, and thus measure their inclination and position angle and compare with those of the outer disks. For the outer disks, we have either collected the inclination and position angles from the literature or performed a fit to determine them (see Section \ref{ssec:outer_disk} and Table \ref{tab:inner_outer_disks}). As a variety of techniques are used to determine the outer disk orientation, we conservatively assume an uncertainty of 5 degrees in inclination and position angle of the outer disk. We compare the inclination and position angle of the (resolved) inner and outer disks in Figure \ref{fig:misalign_i_pa}. In 8 of 14 resolved disks, a significant ($>2\sigma$ probability) misalignment in either position angle or inclination is found. For the 6 other disks, the error bars are such that alignment cannot be confirmed.

Inner dust disk orientations have also been measured through VLTI observations with PIONIER and GRAVITY in H-band and K-band respectively for a number of Herbig stars \citep{Lazareff2017,Perraut2019}. The latter have higher S/N due to the higher contrast with the stellar photosphere. Comparing their samples with our inner disk resolved detections we find only limited overlap: HD100546, HD169142, AB~Aur and MWC758, where the latter two were only part of the PIONIER survey \citep{Lazareff2017}. Our derived orientations for HD100546 and HD169142 have very large error bars so they are consistent with the GRAVITY results from \citet{Perraut2019}. For AB~Aur and MWC~758 we find very different values than \citet{Lazareff2017}, suggesting that the inner disk might be warped  as the millimeter emission is primarily tracing the outer part and the NIR primarily the inner part.

A misalignment between inner and outer disk can be the result of the presence of a companion, which breaks and warps the disk under low viscosity conditions \citep[e.g.][]{Lodato2013,Facchini2018warps,OwenLai2017} and its detection could be interpreted as indirect evidence for the presence of a massive companion.

In Table \ref{tab:misalignment}, we summarize the misalignment found for our detected inner disks, and list other indirect signatures of a misaligned inner disk where known for our sample. These indirect signatures include shadows on the outer disk seen in scattered light, deviations from a Keplerian velocity pattern in the CO gas motion, and a "dipper" host star which shows aperiodic/quasiperiodic dimming episodes that may be due to extinction by an edge-on inner disk (in contrast to a more face-on outer disk). As several of our disks have poorly constrained inclination or position angle and not every disk has been surveyed for indirect signatures of misalignment, we only indicate with a 'Y' if a signature of misalignment is known in the literature. 

Of the 8 objects in our sample with an indirect signature of a misaligned inner disk, we detect an inner disk in 5 cases (AA Tau, HD100453, HD100546, HD142527, MWC 758), although the inclination and position angle are poorly constrained except for MWC 758, which shows a robust misalignment in the mm inner disk. In DoAr44 and J1604.3-2130 where we do not detect an inner disk but indirect signatures of misalignment are known, we have upper limits on the dust mass of $0.025 M_\earth$ and $0.013 M_\earth$ respectively. This may simply be due to a lack of sensitivity in these observations, as other disks in our sample are detected with masses as low as $0.003 M_\earth$ (e.g. MWC 758). Furthermore, a recent study of the light curve of the J1604.3-2130 dipper \citep{Sicilia-Aguilar2019} found that only $2\times10^{-3} - 5\times10^{-3} M_\text{Ceres}$ ($M_\text{Ceres} \approx 1.5\times10^{-4} M_\earth$) was required to reproduce the observed eclipses, well below the detection limits in the ALMA observations. Overall, a misaligned disk may be present in 12 of 18 of our detected disks, and none of our resolved objects can be confirmed to be not misaligned. This suggests that misaligned disks at a few AU scales are common in transition disks, and a plausible way to explain outer disk shadows and CO warps. Previous ALMA observations of dipper star outer disks \citep{Ansdell2016dipper, Ansdell2019} have shown that the outer disks of dipper stars have an isotropic inclination distribution, requiring frequent misalignment of the inner disk to bring it close to edge-on, or mechanisms to reproduce the dipper behaviour which do not require an edge-on disk. The high frequency of misaligned inner and outer disks in our sample thus suggests that dipper star light curves are best modelled with mechanisms requiring an edge-on inner disk. We emphasize that dippers are a special case of misalignment, i.e. where the inner disk ends up in an edge-on orientation obscuring the star, whereas a much larger fraction of possible misalignments would not result in obscurations of the star and dipper behaviour.

One other explanation for apparent misalignment of the inner disk in the millimeter is that we are not measuring inner disk dust emission, but free-free emission from a jet which is perpendicular to the protoplanetary disk. This has been proposed to explain the emission morphology of JVLA 3.3 cm emission in the inner part of the AB Aur disk \citep{Rodriguez2014}. High angular resolution observations at centimeter wavelengths are required to reliably distinguish between jets and inner dust disks. However, indirect evidence of misaligned inner disks, such as shadows, dippers and warps in the kinematics which have been found in many of our targets support the interpretation of misaligned inner disks over jets in several disks already.

\begin{deluxetable*}{cccccccccccc}
\label{tab:misalignment}
\tablecaption{Inner and outer disk (mis)alignment.}
\tablehead{
\colhead{Name} & \colhead{Inner Disk?} & \multicolumn{2}{c}{Inner i, PA} & \multicolumn{2}{c}{Outer i, PA} & \colhead{Mis. i} & \colhead{Mis. PA} & \colhead{Shadows} & \colhead{CO Warp} & \colhead{Dipper} & \colhead{Ref.}\\ \colhead{ } & \colhead{ } & \colhead{($^\circ$)} & \colhead{($^\circ$)} & \colhead{($^\circ$)} & \colhead{($^\circ$)} & \colhead{ } & \colhead{ } & \colhead{ } & \colhead{ } & \colhead{ }& \colhead{ }}
\startdata
AATau & Y & $55 \pm 25$ & $26 \pm 88$ & 59 & 93 & - & - & - & Y & Y& -,6,12 \\
ABAur & Y & $55 \pm 8$ & $36 \pm 12$ & 23 & 36 & Y & - & - & - & - \\
CIDA9 & N & - & - & 46 & 103 & - & - & - & - & - \\
CQTau & N & - & - & 35 & 55 & - & - & - & - & - \\
CSCha & N & - & - & 8 & 161 & - & - & - & - & - \\
DMTau & Y & $41 \pm 4$ & $141 \pm 7$ & 35 & 158 & - & - & - & - & - \\
DoAr44 & N & - & - & 20 & 30 & - & - & Y & - & - & 1,-,-\\
GGTau AA/Ab & Y & $57 \pm 3$ & $25 \pm 3$ & 36 & 98 & Y & Y & Y & - & - & 2,-,-\\
GMAur & Y & - & - & 53 & 56 & - & - & - & - & - \\
HD100453 & Y & - & - & 30 & 149 & - & - & Y & Y & - & 3,7,-\\
HD100546 & Y & $26 \pm 63$ & $169 \pm 83$ & 42 & 139 & - & - & - & Y & - & -,8,-\\
HD135344B & N & - & - & 12 & 62 & - & - & - & - & - \\
HD142527 & Y & $41 \pm 24$ & $41 \pm 24$ & 27 & 25 & - & - & Y & Y & - & 4,9,- \\
HD169142 & Y & $35 \pm 37$ & $68 \pm 84$ & 12 & 5 & - & - & - & - & - \\
HD34282 & N & - & - & 59 & 117 & - & - & - & - & - \\
HD97048 & Y & - & - & 41 & 4 & - & - & - & - & - \\
HPCha & Y & $37 \pm 25$ & $45 \pm 43$ & 37 & 162 & - & Y & - & - & - \\
IPTau & N & - & - & 45 & 173 & - & - & - & - & - \\
IRS48 & N & - & - & 50 & 100 & - & - & - & - & - \\
J1604.3-2130 & N & - & - & 6 & 80 & - & - & Y & Y & Y &5,10,13\\
LkCa15 & N & - & - & 55 & 60 & - & - & - & - & - \\
MHO2 & N & - & - & 38 & 120 & - & - & - & - & - \\
MWC 758 & Y & $50 \pm 6$ & $7 \pm 9$ & 21 & 62 & Y & Y & - & Y & - \\
PDS70 & Y & $29 \pm 9$ & $152 \pm 22$ & 52 & 157 & Y & - & - & - & - \\
PDS99 & N & - & - & 55 & 107 & - & - & - & - & - \\
RXJ1842.9-3532 & N & - & - & 32 & 30 & - & - & - & - & - \\
RXJ1852.3-3700 & N & - & - & 30 & 124 & - & - & - & - & - \\
RYLup & N & - & - & 67 & 109 & - & - & - & - & - \\
RYTau & N & - & - & 65 & 23 & - & - & - & - & - \\
SR21 & N & - & - & 16 & 14 & - & - & - & - & - \\
SR24S & Y & $56 \pm 5$ & $77 \pm 7$ & 46 & 23 & - & Y & - & - & - \\
Sz91 & N & - & - & 45 & 17 & - & - & - & - & - \\
Tcha & Y & - & - & 73 & 113 & - & - & - & - & - \\
TWHya & Y & $55 \pm 1$ & $72 \pm 1$ & 7 & 155 & Y & Y & - & - & - \\
UXTauA & N & - & - & 40 & 167 & - & - & - & - & - \\
V1247Ori & N & - & - & 30 & 115 & - & - & - & - & - \\
V4046Sgr & Y & $59 \pm 5$ & $76 \pm 6$ & 34 & 67 & Y & - & - & - & - \\
WSB60 & Y & $27 \pm 8$ & $2 \pm 50$ & 28 & 172 & - & - & - & - & -
\enddata
\tablecomments{
Reference order: Shadows, CO Warp, dipper. The misalignment between the inner and outer disk was derived in this work.
{\bf Refs.}
1) \citet{Casassus2018}; % DoAr 44
2 \citet{Brauer2019}; %% GG Tau
3) \citet{Benisty2017}; % HD 100453
4) \citet{Marino2015}; % HD142527
5) \citet{Pinilla2018}; % J1604
CO Warps:
6) \citet{Loomis2017}; % AA Tau
7) \citet{vanderPlas2019}; % HD100453
8) \citet{Walsh2017}; % HD 100546
9) \citet{Casassus2013}; % HD 142527
10) \citet{Mayama2018}; % J1604
11) \citet{Boehler2018}; % MWC 758
Dippers:
12) \citet{Bouvier1999, Bouvier2007}; % AA Tau
13) \citet{Sicilia-Aguilar2019}. % J1604
}
\end{deluxetable*}

\subsection{NIR vs millimeter emission}
\label{ssec:inner_nir}
\begin{figure}[htb]
    \centering
    \includegraphics[scale=1.0]{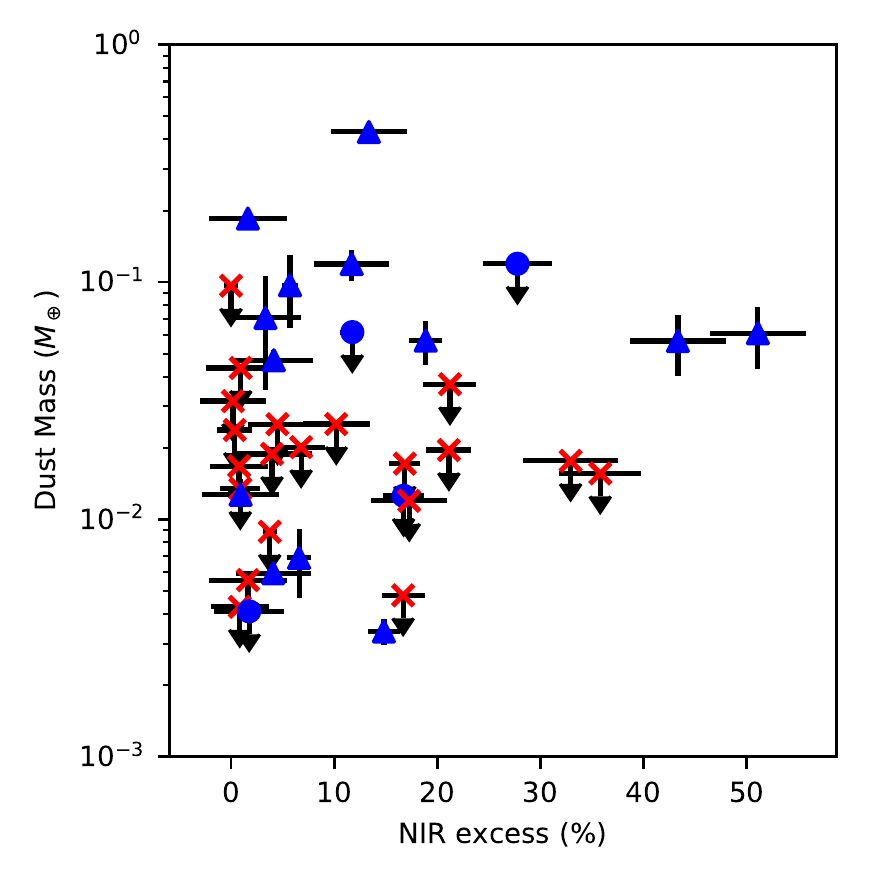}
    \caption{Inner disk dust mass vs NIR excess from the SED defined in Section \ref{ssec:nir_excess_calc}. Blue symbols show detections (triangles resolved, circles unresolved) and red symbols show non-detections. No correlation was found between these two parameters ($r_\text{corr}=0.30 \pm 0.18$), implying that the NIR excess is not a good measure of the presence of mm dust in the inner disk.}
    \label{fig:dust_mass_nir}
\end{figure}

Traditionally, the presence of an inner disk has been assessed using the NIR excess in the SED. A plot of our derived inner disk dust mass against percentage NIR excess is shown in Figure \ref{fig:dust_mass_nir}. No clear correlation is apparent from the data. 
Using the linear regression procedure of \cite{Kelly2007}, which takes into account upper limits and intrinsic scatter in the data, we find no significant correlation between these properties: the correlation coefficient $r_\text{corr}=0.30 \pm 0.18$. The NIR excess (and similarly, the PTD/TD classification by the $\Delta K$ value) is thus not necessarily a reliable measure of the presence of a inner mm-dust disk. This can be understood as the two wavelengths are tracing different regimes: whereas the NIR emission is primarily originating from micron-sized grains at the hot inner dust wall, the millimeter emission is dominated by the millimeter-sized dust grains in the outer part of the inner disk, which can be several au in radius. The lack of correlation suggests that many of the inner mm-dust disks are actually inner rings, with an inner radius well beyond the sublimation radius (and thus no longer detected in the NIR). This applies in particular to those systems with a detected inner disk in the millimeter but low NIR excess  ($< 5\%$): DM Tau, V4046 Sgr, GM Aur, PDS 70, and TW Hya.

The lack of correlation between dust mass and NIR excess also appears to contradict some of the results by \citet{Banzatti2018} on their interpretation of rovibrational lines of Herbig disks. They find that low NIR excess correlates with large inner CO radii, sub-solar metallicity and higher CO excitation levels ($v2/v1$), which they interpret as full clearing of the inner dust disk within the planet's orbit. They also identify a second category of disks with high NIR excess, smaller CO radii, solar metallicity and lower CO excitation levels, interpreted as a massive inner disk inside the planet's orbit. Based on these results, they propose a dichotomy in the presence or absence of inner disk. Comparing their sample with our observed Herbig stars, we have 9 overlapping targets. We detect millimeter-dust inner disks in 4 out of 5 high NIR disks, consistent with massive inner disks. However, we also detect mm-dust inner disks in 3 out of 4 of the low NIR disks, with outer radii $R_\text{inner}$ well within their derived inner CO radii ($R_{CO}$). The derived dust masses for each category are not showing a dichotomy. Whereas the lower NIR excess implies that the $R_\text{in}$ of the inner disk is beyond the sublimation radius (potentially caused by another close-in planet), our results contradict the proposed scenario where no dust is present inside the $R_{CO}$ line. 

%\newpage
\subsection{Dust depletion in the inner disk}
\label{ssec:inner_dust}
\begin{figure}[htb]
    \centering
    \includegraphics[]{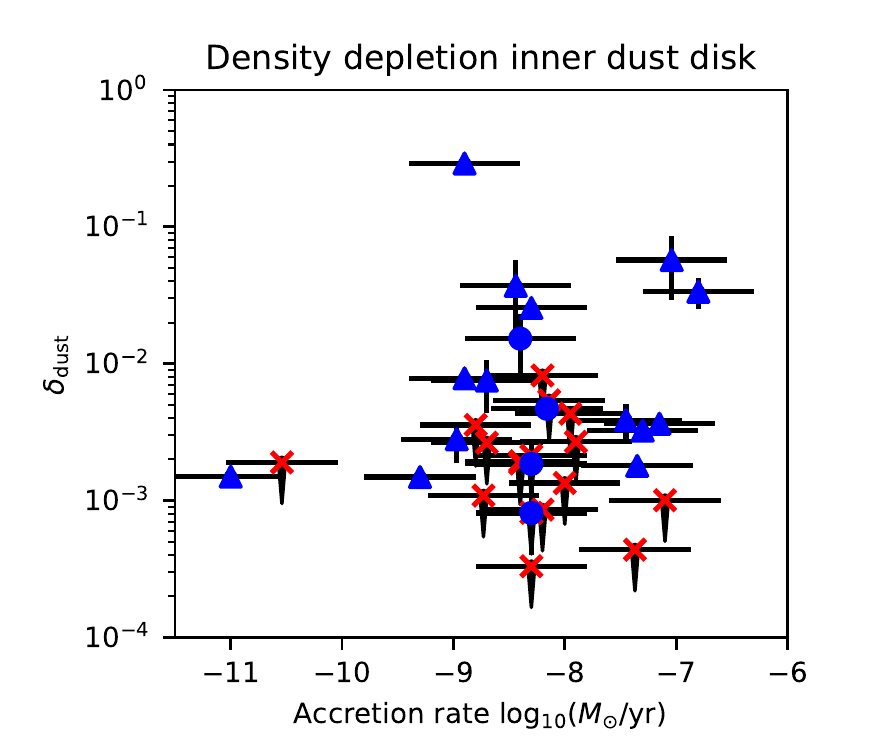}
    \caption{Fractional dust depletion of the inner disk relative to the outer disk vs the accretion rate. 0.5 dex of error is assumed for the accretion rate. Blue symbols show detections (triangles resolved, circles unresolved) and red symbols show non-detections. All inner disks show a clear depletion of dust with respect to the outer disk. The data are not anti-correlated ($r_\text{corr}=0.052 \pm 0.54$) of $\delta_\text{dust}$ with accretion rate, consistent with the millimeter-grains being decoupled from the accreting gas.}
    \label{fig:ddust}
\end{figure}

From the dust surface density profiles in Figure \ref{fig:surfdensity} it is clear that the dust surface density drops by more than an order of magnitude in the inner disk compared to the outer disk. In Figure \ref{fig:ddust}, we compare the fractional drop $\delta_\text{dust}$ from Section \ref{ssec:outer_disk} across the sample.

All of the disks in our sample have a $\delta_\text{dust} < 1$, with a median of $\delta_\text{dust}\approx 10^{-2}$, indicating significant dust depletion relative to the outer disk. As our sample covers a wide range of disk properties, this implies that the mechanism responsible for the dust depletion operates rapidly, on timescales shorter than the lifetimes of the youngest disks. 
 
 Furthermore, the value of $\delta_\text{dust}$ remains constant across a range of accretion rates: we see no anti-correlation ($r_\text{corr}=0.052 \pm 0.54$) of $\delta_\text{dust}$ with accretion rate, consistent with the millimeter-grains being decoupled from the accreting gas.

These two effects (rapid dust depletion and decoupling of dust from the gas) can be explained by rapid initial radial drift in the inner disk, when separated from the outer disk. In a smooth, continuous disk, millimeter grains move at sub-keplerian velocities, and consequently feel a headwind that causes them to rapidly drift toward the star and sublimate \citep{Weidenschilling1977}. However, both embedded planets and instabilities (e.g. dead zones) can create pressure bumps in the outer part of the disk which limits the inward drift of millimeter grains to the location of the pressure bump \citep{Pinilla2012b}, provided a low value of the viscosity parameter $\alpha<= 10^{-3}$ . This results in a pile-up of mm grains at the pressure bump, and a rapid inward radial drift of millimeter grains in the inner disk which can not be replenished due to a cutoff of the flow of dust grains from the outer disk. Smaller dust grains may continue to flow through the gap after the outer pressure bump has been established, and grow to larger sizes again in the inner disk through coagulation, but for a sufficiently massive planet even the smallest dust grains cannot flow inwards any more. Inward drift results in depletion of mm-dust of the inner disk as dust grains may either sublimate at the sublimation radius or grow to large pebbles and rocks. 

\subsection{Inner disk size}
\label{ssec:innerdisksize}

\begin{figure}[htb]
    \centering
    \includegraphics[scale=1.0]{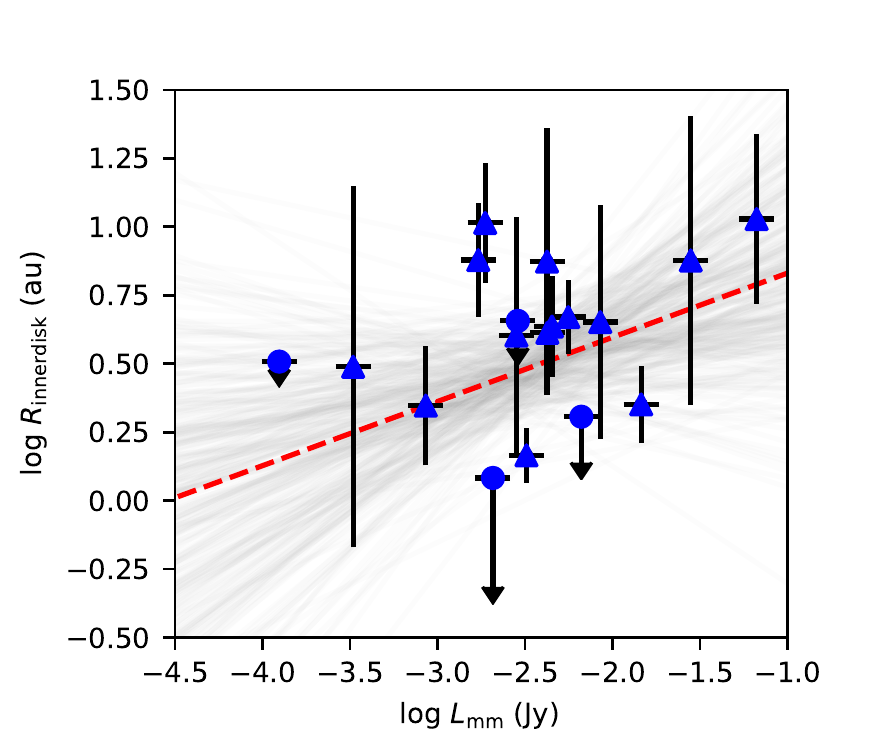}
    \caption{Millimeter continuum size-luminosity relationship for the inner dust disk, as derived from our analysis. Inner disk dust size is defined as $R_\text{inner}$, and the $L_\text{mm}$ is the flux scaled to a distance of 140 pc and to a frequency of 340 GHz (assuming a spectral index $\alpha_{mm}$=2.5), similar to \citet{Andrews2018size}. The red line shows the scaling relation from the best fit linear regression analysis, while the gray lines show the variation in fits derived from the Markov Chain Monte Carlo samples of the posterior probability distribution. The best fit line has a correlation coefficient $r_\text{corr}=0.5 \pm 0.3$ and parameter values (see text) consistent with the correlation found for full protoplanetary disks \citep{Andrews2018size}.}. 
    \label{fig:innerdisksize}
\end{figure}

Protoplanetary dust disks have been found to follow a continuum size - 
luminosity relation, demonstrating that the amount of emission scales linearly with the emitting surface area \citep{Tripathi2017,Andrews2018size}. This correlation could be reproduced by dust evolution models in the regime dominated by radial drift rather than fragmentation \citep{Rosotti2018}. For our sample of inner disks, we check if the same correlation exists. 

In Figure \ref{fig:innerdisksize} we present the relationship between continuum size (represented by $R_\text{inner}$) and the millimeter luminosity, which is the millimeter flux of the inner dust disk, scaled to a distance of 140 pc and a frequency of 340 GHz for a spectral index $\alpha_{mm}$=2.5 (similar to \citet{Andrews2018size}). We only include the detections in this plot, although several values of $R_\text{inner}$ are upper limits when the inner disk is unresolved. The linear regression procedure mentioned before shows a moderate correlation with $r_\text{corr}=0.5\pm0.3$, with a linear relation:
\begin{equation}
\log R_\text{inner} = A + B \log L_\text{mm}    
\end{equation}
with best-fit parameters $A= 1.2\pm0.6$ and $B=0.3\pm0.2$. Although the error bars are much larger, this is only slightly offset but with a slope that is consistent with the results of \citet{Andrews2018size} for a sample of more than 100 protoplanetary disks, who find $A=2.1\pm0.05$ and $B=0.5\pm0.05$ for this relation. This implies that inner dust disks as isolated systems are susceptible to radial drift and follow a similar morphology as full disks.

\begin{figure}[htb]
    \centering
    \includegraphics[scale=1.0]{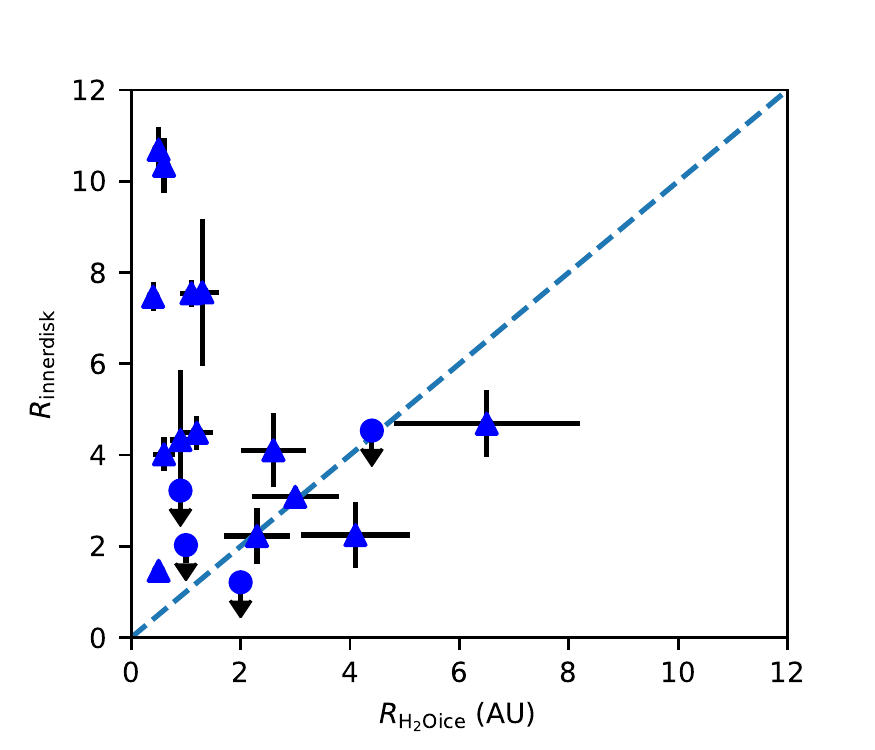}
    \caption{Relationship between the outer radius of the inner disk $R_\text{inner}$ and the location of the water snowline, taken at 140$\pm$15 K. The dashed line indicates the one-on-one relationship, but most targets do not follow the correlation: $R_\text{inner}$ is well outside the snowline.}. 
    \label{fig:snowline}
\end{figure}

Second, we compare the outer radius of the inner dust disk with the location of the water snowline in each system in Figure \ref{fig:snowline}. The water snowline is defined as the radius where the temperature drops below the freeze-out temperature of H$_2$O, which is taken as 140$\pm$15 K \citep{Zhang2015}. The snowline radius is computed from Equation \ref{eqn:temperature} using the stellar luminosity. For the majority of the disks the outer radius of the dust disk is well outside the snowline. For a handful of the targets the radius is similar (within 1 au). 

The snowline is thought to affect the dynamics of the dust particles because grains without ice mantles are expected to stick less efficiently and therefore the fragmentation velocity (velocity threshold from destructive collisions) decreases inside the snowline \citep[e.g.][]{Birnstiel2010}. Right outside the snowline, the dust growth is still set by the radial drift. \citet{Pinilla2016traffic} ran a series of dust evolution simulations for a range of companion masses to quantify the dust growth in the inner disk. They found that for a low-mass companion (1 M$_{\rm Jup}$), the inner mm-dust disk remains large even after 5 Myr as dust in the inner disk gets replenished through the gap. However, for a high-mass companion (5 M$_{\rm Jup}$) the inner mm-dust disk eventually shrinks due to radial drift as no dust flows through the gap. In the latter case, the micron-sized grains inside the snow line also disappear.

Since we find mm-dust well beyond the snowline for several of our targets, this implies that either the dust flow has only recently been cut off and radial drift has not fully shrunk the inner dust disk yet, or the mass of the companion is low enough to allow replenishment of the inner dust disk. It is also possible that the snowline location cannot be described by a simple power-law for the temperature relation. More detailed dust evolution modeling of the effect of the snowline and the possible companion masses is required to fully test this phenomenon.

\subsection{Gas content of the inner disks}
\label{ssec:inner_gas}

\begin{figure}[htb]
    \centering
    \includegraphics[scale=1.0]{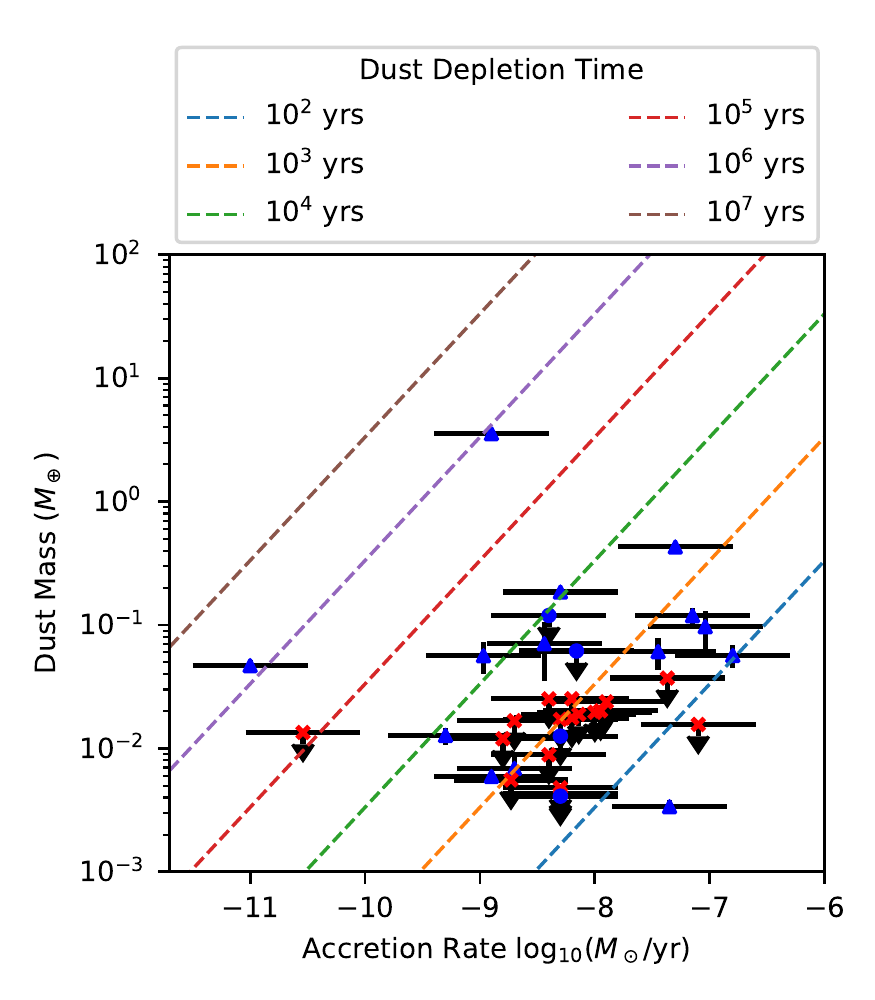}
    \caption{Inner disk dust mass vs accretion rate. 0.5 dex of error is assumed for the accretion rate. An ISM-like gas to dust ratio of 100 is assumed for the inner disk, the validity of which is discussed in \ref{ssec:inner_gas}. No correlation appear to be present ($r_\text{corr}=-0.24 \pm 0.26$), which is again consistent with decoupling of the mm grains and accreting gas. Assuming a gas-to-dust ratio of 100, the lifetime of the inner disks is typically less than $10^4$ years, well below the typical lifetime of a protoplanetary disk.} 
    \label{fig:dust_mass_acc}
\end{figure}

The lack of correlation in Figure \ref{fig:ddust} implies a decoupling of the mm grains and gas. We can test this further by assuming instead that there is no decoupling, and comparing the inferred gas content of the inner disk with the accretion rate. Assuming a standard ISM gas to dust ratio, we compare the dust mass and accretion rate in Figure \ref{fig:dust_mass_acc}, and overplot the resulting timescale for the gas and dust content of the inner disk to accrete onto the star. Similar to Figure \ref{fig:ddust}, no correlation appears to be present ($r_\text{corr}=-0.24 \pm 0.26$), which is again consistent with decoupling of the mm grains and accreting gas. Secondly, we find that the lifetimes of most inner disks are relatively short compared to the lifetime of the disk, typically less than $10^4$ yr, assuming an ISM gas-to-dust ratio of 100. 

Given the high occurrence rate of inner disks in our sample, a short lifetime would require that the inner disk is regularly replenished by episodes of enhanced accretion from the outer disk. However, depletion of dust in the inner disk due to radial drift such as suggested in the previous section may cause the gas to dust ratio of the inner disk to be several orders of magnitude larger \citep{Pinilla2012b}, implying proportional longer lifetimes. While we do not have a direct measurement of gas content of the inner disk to assess this scenario, viscous accretion disk theory predicts a link between the gas surface density, accretion rate and viscosity which we can use to estimate the gas content. We assume the following relations:
\begin{align}
\nu\Sigma_g(r) = \frac{\dot{M}}{3\pi} \\
\nu = \alpha c_s H \nonumber \\
H = c_s/\Omega(r) \nonumber \\
c_s = \sqrt{\frac{k_BT(r)}{\mu m_p}} \nonumber \\
\Omega(r) = \sqrt{\frac{GM_*}{r^3}} \nonumber 
\end{align}
with $\nu$ the disk viscosity, $\alpha$ the viscosity parameter, $c_s$ the sound speed, $H$ the vertical height, $\Omega(r)$ the angular velocity, $k_B$ the Boltzmann constant, $\mu\approx2$ the mean molecular weight, $m_p$ the proton mass, $G$ the gravitational constant and $M_*$ the stellar mass. The third equation assumes that the disk is vertically isothermal. These equations result in a relation between the gas surface density and the accretion rate \citep{Manara2014}:
\begin{equation}
\Sigma_g(r)= \frac{\dot{M} 2 m_p}{3\pi\alpha k_BT(r)}\sqrt{\frac{GM_*}{r^3}},
\end{equation}
The gas surface density profile is derived from the dust profile with $\xi_\text{inner}$ and $\xi_\text{outer}$ the gas-to-dust ratio for inner and outer disk, respectively:
\begin{align*}
\Sigma_g(r) &= \xi_\text{outer}\frac{\Sigma_cr_c}{r} & \textrm{for } r>R_\text{cav} \\
&= 0 & \textrm{for } R_\text{inner}<r<R_\text{cav} \\
&= \xi_\text{inner}\delta_\text{dust}\frac{\Sigma_cr_c}{r} & \textrm{for } R_\text{sub}<r<R_\text{inner} 
\end{align*}
while for the temperature profile we use Eqn \ref{eqn:temperature}. This means that we can express the gas-to-dust ratio $\xi_\text{inner}$:
\begin{equation}
\xi_\text{inner} = C\cdot\sqrt{\frac{M_*}{\sqrt{L_*}}}\frac{\dot{M}}{\alpha \delta_\text{dust}\Sigma_cr_c}
\label{eqn:gdr}
\end{equation}
with
\begin{equation}
C=\frac{\sqrt{G} \mu m_p}{3\pi k_B}\sqrt[4]{\frac{8\pi\sigma_B}{\phi}},
\end{equation}
and equivalently, we can express viscosity $\alpha$ as:
\begin{equation}
\alpha = C\cdot\sqrt{\frac{M_*}{\sqrt{L_*}}}\frac{\dot{M}}{\xi_\text{inner}\delta_\text{dust} \Sigma_cr_c}.
\label{eqn:alpha}
\end{equation}

We note that both quantities $\alpha$ and $\xi_\text{inner}$ are independent of $r$ and can be evaluated at any radius, and we can compute either quantity while keeping the other one fixed.
We are thus able to evaluate various assumptions about the values of $\xi_\text{inner}$, $\xi_\text{outer}$, and $\alpha$ to determine whether the gas to dust ratio is enhanced in the inner disk. 

\begin{figure*}[htb]
    \centering
    \includegraphics[]{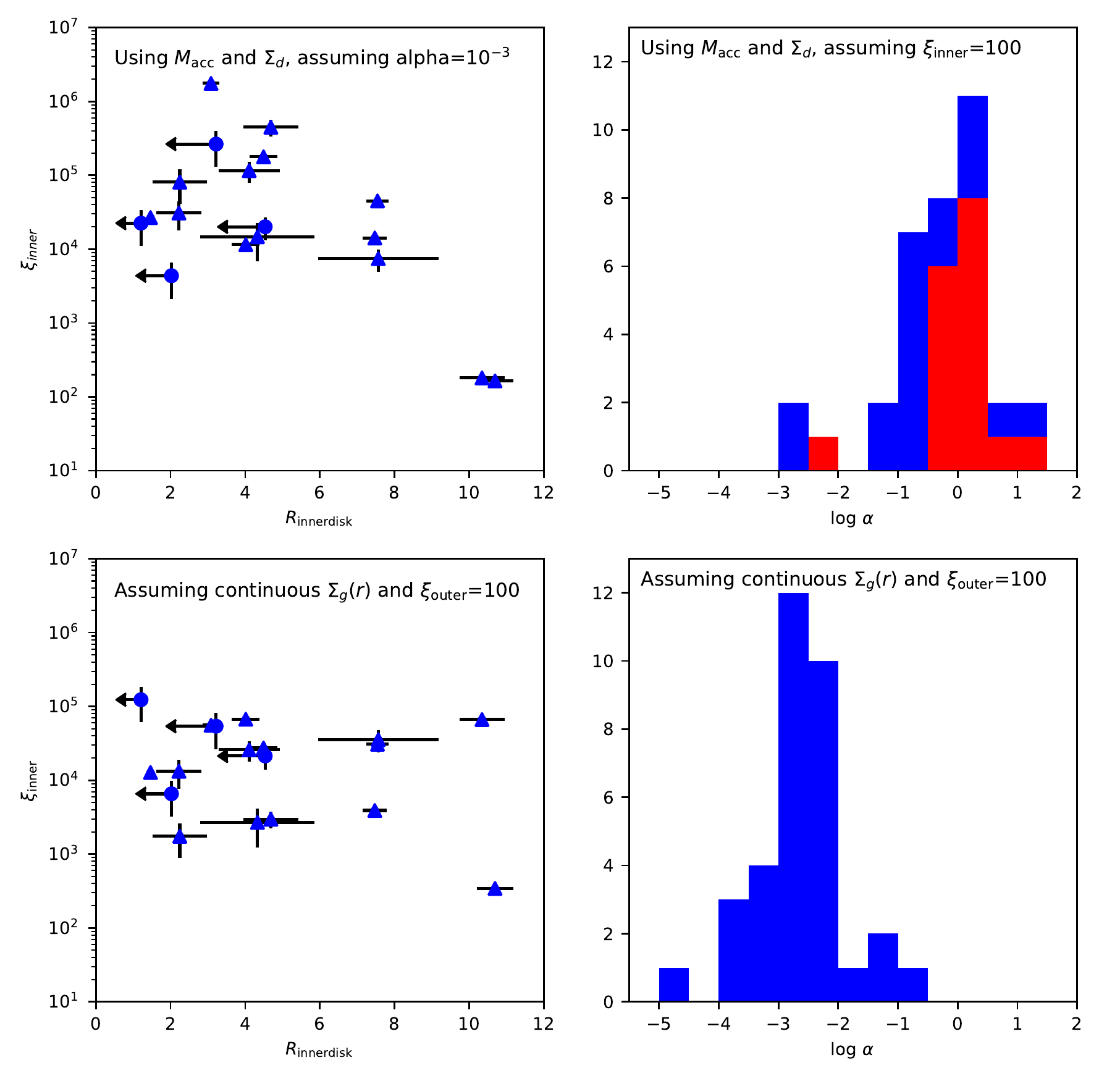}
    \caption{Gas in inner disk vs accretion properties assuming the $\alpha$ disk model. In the top left, the gas-to-dust ratio in the inner disk $\xi_\text{inner}$ is computed from the accretion rate and the derived dust surface density profile, assuming $\alpha=10^{-3}$. In the bottom left, $\xi_\text{inner}$ is computed assuming a continuous surface density profile from the outer disk, with $\xi_\text{outer}$=100, without the need for invoking $\alpha$. Both plots indicate that the gas-to-dust ratio in the inner disk is likely much higher than the ISM ratio of 100, consistent with dust evolution models. The plots on the right show a histogram of the distribution of the $\alpha$ viscosity, assuming $\xi_\text{inner}=100$ (top) and assuming a continuous distribution from the outer disk and $\xi_\text{outer}$=100 (bottom), same as in the lower left. The second scenario results in realistic values of $\alpha$. In the plots, blue triangles are measured and blue circles are unresolved. In the histograms, the red components are lower limits of $\alpha$.} 
    \label{fig:accretion}
\end{figure*}

In the top-left panel of Figure \ref{fig:accretion} we use Equation \ref{eqn:gdr} to calculate $\xi_\text{inner}$ for a low $\alpha$=10$^{-3}$, and find gas-to-dust ratios of $10^4-10^5$ for the majority of the (detected) disks. The non-detections are left out of this plot for clarity, but their limits indicate $\xi_\text{inner}>10^4$ as well. For lower values of $\alpha$, the inferred gas-to-dust ratios would be even larger, while larger values of $\alpha$ are unable to produce a mm dust trap such as observed in the outer disk in these images \citep{Pinilla2012b}.  If the gas to dust ratios are indeed $>100$ times larger than the ISM ratio of 100, the disk lifetimes in Figure \ref{fig:dust_mass_acc} may also be two orders of magnitude larger, and replenishment is no longer needed to explain the high occurrence rates of inner disks in our sample. 

We notice that the two disks with a lower $\xi_\text{inner}$ ratio of $\sim$100 are the largest inner dust disks in the sample, PDS 70 and WSB 60. In these cases, a gap has possibly been opened only recently, and the mm grains have not yet had time to drift toward the star and sublimate. These two disks also lie well above the size-luminosity correlation in Figure \ref{fig:innerdisksize}, implying they are not subject to the radial drift regime which has been used to explain this correlation. The low fraction of 2/38 (assuming all non-detections are smaller and fainter disks) is consistent with rapid initial radial drift. The typical drift time scale for millimeter-sized particles is $\sim10^{5}$ years for typical disk conditions at 1 au \citep{Birnstiel2010}. 

If we instead assume an ISM-like $\xi_\text{inner}$ of 100, we find the $\alpha$ values of most disks in our sample would be $>$~1 (see top right panel of Figure \ref{fig:accretion}), which is completely unphysical for a viscous accretion disk. Alternatively, if we assume a $\xi_\text{outer}$ of 100, and a continuous $\Sigma_g$ profile across the entire disk, $\xi_\text{inner} = \xi_\text{outer}/\delta_\text{dust}$, and $\alpha$ can be computed using $\xi_\text{outer}$ in Equation \ref{eqn:alpha}. The two lower panels of \ref{fig:accretion} show that our $\xi_\text{inner}$ is enhanced by two orders of magnitude or more relative to the outer disk, and suggests $\alpha$ is low for most disks, which is again compatible with the expectations of dust drift. We note that the values of $\xi_\text{inner}$ in this case are also comparable to $\xi_\text{inner}$ values when computed from the accretion rate from the top panel. The ratio between these two values may represent the gas depletion in the inner part of the disk due to accretion, and thus the evolutionary stage, but due to the large uncertainties and assumptions in our derivation of the gas surface we do not attempt to estimate these ages. 

 The previous arguments suggest that the inner disk is dust depleted (or equivalently,  $\xi_\text{inner}$ is enhanced) relative to the outer disk for a low $\alpha$ disk. Gas-to-dust ratios of $\sim10^5$ are also found in the inner disks in dust evolution models with planet-disk interaction due to radial drift \citep{Pinilla2012b}.
 
 However, in recent years the $\alpha$-disk model has become a topic of debate, as the measured turbulence in disks is very low \citep{Flaherty2017,Flaherty2018}, and inclusion of non-ideal MHD effects in disk evolution models has been shown to suppress the viscous spreading \citep[e.g.][]{Bai2016}. This would imply a $\xi_\text{inner}$=100 is possible (but not required), and that the inner disks are may indeed be short-lived structures, as shown in section \ref{ssec:inner_dust_mass}, and thus replenishment of the inner disk by enhanced accretion episodes is required to explain the high occurrence in our sample. If replenishment is occurring, the accreting material must flow at speeds near the free-fall velocity \citep{Rosenfeld2014} to be consistent with the measured low gas surface densities in the gaps of transition disks \citep{vanderMarel2016-isot}.

In order to distinguish between the possibilities, more direct measurements of the gas content of the inner disk are required. Either spatially resolved ALMA observations of CO isotopologues or near infrared observations of rovibrational molecular lines with the \emph{James Webb Space Telescope} or thirty-meter class telescopes may help to constrain the gas content in the inner disk. However, both chemistry and excitation conditions in the inner disk are poorly understood, so interpretation of these data will remain challenging.

\subsection{Consequences for the presence of companions}
\label{ssec:companions}
All of our inner disks show significant depletion in dust surface density relative to the outer disks (Figure \ref{fig:ddust}), and if the inner disks are well described by an $\alpha$-disk model, they must also have a low $\alpha$ and high $\xi_\text{inner}$. These features are expected if a giant ($ > 1 M_\text{Jup}$) planet traps mm grains outside its orbit, while mm grains within the inner disk rapidly drift towards the star and sublimate \citep{Pinilla2012b,Dong2015gaps}. In principle, constraints on the mass and location of the conjectured planets can be placed based on the depth and width of the gap in the gas surface density \citep{DongFung2017}. However, since our study only resolves the dust whose structure is regulated by a combination of planet-disk interaction and radial drift, our results cannot be used to derive properties of embedded planets. Spatially resolved measurements of the gas surface density of molecular tracers from either ALMA or NIR rovibrational lines are required.
  
\begin{figure}[htb]
    \centering
    \includegraphics[width=0.45\textwidth]{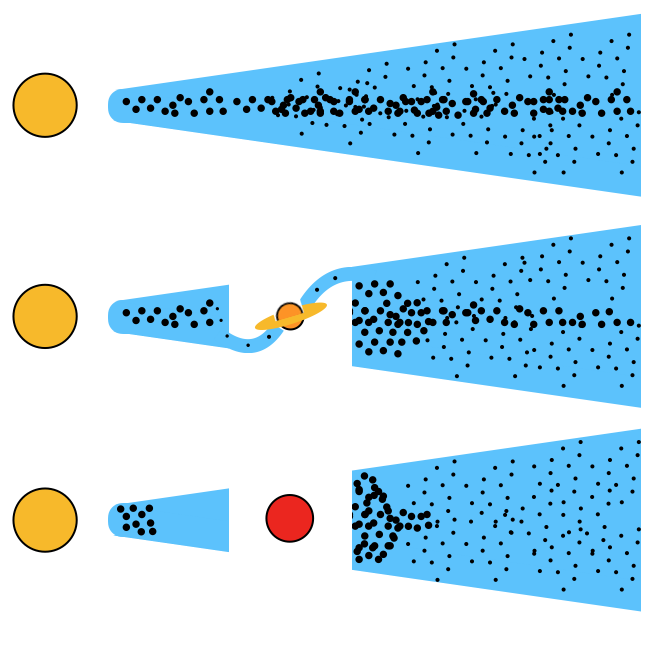}
    \caption{Schematic of proposed gap clearing and depletion mechanism of inner disk. The embedded planet is only detectable when it is still accreting and the inner disk is continuously being replenished and the effects of radial drift in the inner disk are not yet detectable.} 
    \label{fig:protoplanet_cartoon}
\end{figure}

The only disk in our sample confirmed to host an embedded planet \footnote{A second planet in the gap, PDS 70c, has recently been detected in H$\alpha$ emission  \citep{Haffert2019}.} is PDS70 (PDS70b, $\sim 5-9 M_\text{Jup}$, $\sim22$~au) \citep{Keppler2018, Muller2018}. Six other disks in our sample (HD135344B, HD142527, HD97048, MWC 758, UXTauA, and HD100453) have been observed in H$\alpha$ emission with sufficient contrast to redetect PDS 70b, but no planet candidates have been detected \citep{Zurlo2020}. PDS 70b is located just outside the detected inner disk, which is larger ($\sim 10$ au) and has a lower $\xi_\text{inner}$ ($\sim 10^2$) than most other disks in the sample (Figure \ref{fig:accretion}, upper left). We propose the hypothesis that PDS 70b has only recently formed and opened a gap in the disk, such that the supply of dust grains to the inner disk is not yet fully cut off, and the inward drift of mm grains has yet to significantly reduce the size of the inner disk (see schematic representation in Figure \ref{fig:protoplanet_cartoon}). This could also explain why a planet has been detected in the PDS 70 disk, but not in other transition disks where searches have been conducted with similar sensitivity \citep[e.g.][]{Maire2017, Langlois2018, Ligi2018}: PDS 70b is young and actively accreting, and thus it is significantly brighter and has a visible circumplanetary disk.   

The WSB 60 disk in our sample is also an outlier with a large $\sim10$ au inner disk and inferred $\xi_\text{inner}$ of $10^2$. WSB60 is thus similar to PDS~70 and an excellent candidate for direct imaging searches. However, WSB 60 is optically faint %(QUANTIFY, REFERENCE? NM: this is really not necessary, optical photometry has been known for decasdes
and considering the dust cavity radius of $\sim$30 au the planet is likely much further in than for PDS~70, which may make direct imaging challenging.

This hypothesis also suggests the possibility of an evolutionary connection between transition disks and some of the observed ring disks \citep[e.g. DSHARP,][]{Andrews2018dsharp,vanderMarel2019}, where several disks show a bright inner disk surrounded by one or more wide dust gaps, e.g. HD143006, SR4 or Elias~24. Perhaps in these systems a planet was only recently formed, and the inner dust disk has not started to deplete as material continues to flow through the gap. This would lead to a natural evolutionary connection between ring disks and transition disks such as previously suggested \citep{vanderMarel2018}.

Other than the consequences for wide orbit companions, our results can also constrain the presence of close-in companions at a few au orbital distances. Several of the disks in our sample have a low NIR excess but a detected mm disk (Section \ref{ssec:inner_nir}, DM Tau, V4046 Sgr, GM Aur, PDS 70, and TW Hya.), suggesting an inner ring rather than an inner disk. This may be caused by clearing of the inner disk by an unseen companion planet or star in the very inner part of the disk. Of these disks, V4046 Sgr is known to be a spectroscopic equal-mass binary with a separation of only 0.045 au \citep{Stempels2004,Rosenfeld2013}, while the others may contain yet undiscovered planets or low-mass stars within a few au. 

Two of the detected inner disks in our sample are likely circumprimary disks within a multiple star system. HD142527 ($R_\text{inner} 4.1 \pm 0.8$ au) is known to have a binary M-star (mass ratio $q\approx$0.2) companion at a highly eccentric orbit, with semi-major axes ranging from 26 to 50 au \citep{Lacour2016}. GG Tau A is in fact a triple system with a separation of 0.25" or 35 au between the Aa and Ab components \citep{White1999} where Ab consists of Ab1 and Ab2 at a separation of 5 au \citep{diFolco2014}. The B companion is located $\sim$10'' south \citep{Leinert1993} and is less relevant for the study of the disk around GG Tau A. Our detected inner disk is located around the Aa component with a radius of $\sim 7.5$~au, consistent with truncation by the Ab companion, and an inclination angle of $\sim 57^\circ$, consistent with shadowing observed on the outer disk \citep{Brauer2019}.

\section{Conclusions}
\label{sec:conclusions}

In this study, we have collected ALMA archival continuum observations of 38 transition disks with resolved dust cavities, for which we have measured the size and dust mass of the inner disk, or placed upper limits. We have also constructed dust surface density profiles of the inner and outer disk from the ALMA data, measured the NIR excess from archival photometry, and compared the results with stellar properties and accretion rates from the literature. Our main findings are as follows:

\begin{itemize}
    \item At least 18 of 38 transition disks in our sample host an inner dust disk, 14 of which are resolved with a mean radius of $\sim 5$~au. As our sensitivity is extremely non-uniform across the sample, this is likely an underestimate of the occurrence rate of inner disks. 
    \item Of our 14 resolved inner disks, 8 have misalignments in inclination and position angle compared with the outer disk. The other 4 have detected inner disks where CO warps, outer disk shadows, or a dipper host star suggest a misaligned inner disk, which is an indicator of massive giant companions.    
    \item The NIR excess in the SED is uncorrelated with the dust mass of the inner disk, suggesting that the NIR excess is not a reliable measure of the presence of a mm dust inner disk, and that some of our dust disks are in fact dust rings.
    \item The dust surface density of all our inner disks is depleted relative to the outer disk, with a median depletion of $\sim 10^{-2}$. As our sample spans a wide range of disk properties, this suggests the mechanism responsible for depletion operates on timescales shorter than the lifetimes of the youngest disks. 
    \item The continuum size-luminosity correlation found in protoplanetary disks is reproduced for the inner dust disks in our sample, indicating that the dust is in the regime dominated by radial drift. 
    \item If our inner disks are well described by a viscous $\alpha$ disk model, we find a low $\alpha$ ($\sim 10^{-3}$) and high gas-to-dust ratio ($10^4-10^5$) for the inner disk, which implies trapping of mm grains in the outer disk and depletion of mm grains by radial drift in the inner disk.
    \item Alternatively, if the inner disk is poorly described by viscous disk theory, the inner disk gas to dust ratio may be low, implying that lifetimes of the inner disk are short ($< 10^4$~yr), and periods of enhanced accretion from the outer disk with high speed radial flows of gas through the gap are required to explain the high frequency of inner disks. 
    \item In the $\alpha$-disk scenario, the depletion and dust trapping seen in the outer and inner disks respectively is well explained by dust evolution models of planet-disk interaction models involving embedded giant planets.
    \item The only disk with a confirmed planet in our sample, PDS 70, has an inner disk with a large size and an implied low gas-to-dust ratio. This may be explained if the gap has only been opened by the planet recently. We propose a hypothesis that PDS 70 has been the only embedded planet detection to date due to its recent formation, implying material flowing through the gap, active accretion and a bright circumplanetary disk. 

\end{itemize}

\section{Acknowledgements}
We thank the anonymous referee for a careful review and Doug Johnstone, Ruobing Dong, Jeffrey Fung, Carlo Manara, Paola Pinilla, Giovanni Rosotti and Takayuki Muto for useful discussions. 
N.M. acknowledges support from the Banting Postdoctoral Fellowships program, administered by the Government of Canada. This research was enabled in part by support provided by Compute Canada.

The National Radio Astronomy Observatory is a facility of the National Science Foundation
operated under agreement by the Associated Universities, Inc. 

ALMA is a partnership of ESO
(representing its member states), NSF (USA) and NINS (Japan), together with NRC (Canada) and NSC and
ASIAA (Taiwan) and KASI (Republic of Korea), in cooperation with the Republic of Chile. The Joint
ALMA Observatory is operated by ESO, AUI/ NRAO and NAOJ. This paper makes use of the following ALMA
data: 
2011.1.00399.S, 2012.1.00129.S, 2012.1.00158.S, 2012.1.00631.S, 2012.1.00761.S, 2013.1.00100.S, 2013.1.00198.S, 2013.1.00498.S, 2013.1.01070.S, 2015.1.00192.S, 2015.1.00224.S, 2015.1.00678.S, 2015.1.00686.S, 2015.1.00888.S, 2015.1.00889.S, 2015.1.00979.S, 2015.1.00986.S, 2015.1.01083.S, 2015.1.01207.S, 2015.1.01301.S, 2016.1.00344.S, 2016.1.00629.S, 2016.1.00826.S, 2016.1.01042.S, 2016.1.01164.S, 2016.1.01205.S, 2016.A.00026.S, 2017.1.00449.S, 2017.1.00492.S, 2017.1.00884.S, 2017.1.00969.S, 2017.1.01151.S, 2017.1.01167.S, 2017.1.01404.S, 2017.1.01424.S, 2017.1.01460.S, 2017.1.01578.S, 2017.A.00006.S.

This work makes use of the following software: The Common Astronomy Software Applications (CASA) 
package \citep{casa2007}, Python version 
2.7, astropy \citep{astropy2013}, and matplotlib \citep{matplotlib2007}.   

\bibliography{inner_disks}{}
\bibliographystyle{aasjournal}

\appendix

\section{Inner Disk Coordinates}
\begin{deluxetable}{ccc}[h]
\label{tab:inner_disk_coords}
\tablecaption{Detected Inner Disk coordinates.}
\tablehead{\colhead{Name} & \colhead{RA} & \colhead{dec}\\ \colhead{ } & \colhead{(ICRS J2000)} & \colhead{(ICRS J2000)}}
\startdata
AATau & 04:34:55.4277 & +24:28:52.668 \\
ABAur & 04:55:45.8515 & +30:33:03.895 \\
DMTau & 04:33:48.7486 & +18:10:09.650 \\
GGTau AA/Ab & 04:32:30.3663 & +17:31:40.198 \\
GMAur & 04:55:10.987 & +30:21:58.943 \\
HD100453 & 11:33:05.502 & -54:19:28.64 \\
HD100546 & 11:33:25.306 & -70:11:41.232 \\
HD142527 & 15:56:41.870 & -42:19:23.703 \\
HD169142 & 18:24:29.7759 & -29:46:49.999 \\
HD97048 & 11:08:03.186 & -77:39:17.474 \\
HPCha & 11:08:15.366 & -77:33:53.432 \\
MWC 758 & 05:30:27.534 & +25:19:56.600 \\
PDS70 & 14:08:10.107 & -41:23:52.995 \\
SR24S & 16:26:58.506 & -24:45:37.278 \\
Tcha & 11:57:13.2868 & -79:21:31.668 \\
TWHya & 11:01:51.8183 & -34:42:17.238 \\
V4046Sgr & 18:14:10.486 & -32:47:35.479 \\
WSB60 & 16:28:16.5040 & -24:36:58.527
\enddata
\end{deluxetable}

\section{Orientation outer disk}
For 7 targets (CS~Cha, HP~Cha, MHO~2, PDS~99, RXJ1842.9-3532, UX~TauA and WSB~60), no previous fitting of the outer disk orientation is available from the literature. In order to determine the orientation of the outer disk, we perform a simple fitting procedure for each image, where the intensity model is a Gaussian ring with an inclination $i$ and a position angle PA:
\begin{equation}
    I(r) = I_0 e^{(r-r_c)^2/(2r_w^2)}
\end{equation}
with the center of the ring $r_c$ and a width $r_w$. The intensity model is convolved with the beam and subtracted to check the residual. After an initial fit by eye, a $\chi^2$ minimization with steps of 0.005" in radial direction and 1$^{\circ}$ results in the best fit values reported in Table \ref{tbl:outerdiskfit}. The residual images often still contain remnant emission, which are the result of additional structure in the disk that is not included in this model. However, for the purpose of this work (the orientation of the outer disk to within a few degrees) it is sufficient. 

\begin{table}[!ht]
    \centering
    \caption{Outer disk ring best fit parameters .}
    \label{tbl:outerdiskfit}
    \begin{tabular}{ccccc}
        \hline
        \hline
         Name&$r_c$&$r_w$&$PA$&$i$  \\
         &(")&(")&($^{\circ}$)&($^{\circ}$)\\
         \hline
CSCha	&	0.225	&	0.055	&	161	&	8	\\
HPCha	&	0.28	&	0.04	&	162	&	37	\\
MHO2	&	0.245	&	0.05	&	120	&	38	\\
PDS99	&	0.42	&	0.08	&	107	&	55	\\
RXJ1842	&	0.27	&	0.07	&	30	&	32	\\
UXTauA	&	0.25	&	0.03	&	167	&	40	\\
WSB60	&	0.25	&	0.08	&	172	&	28	\\
\hline
    \end{tabular}
\end{table}

\begin{figure}
    \centering
    \includegraphics[scale=0.9]{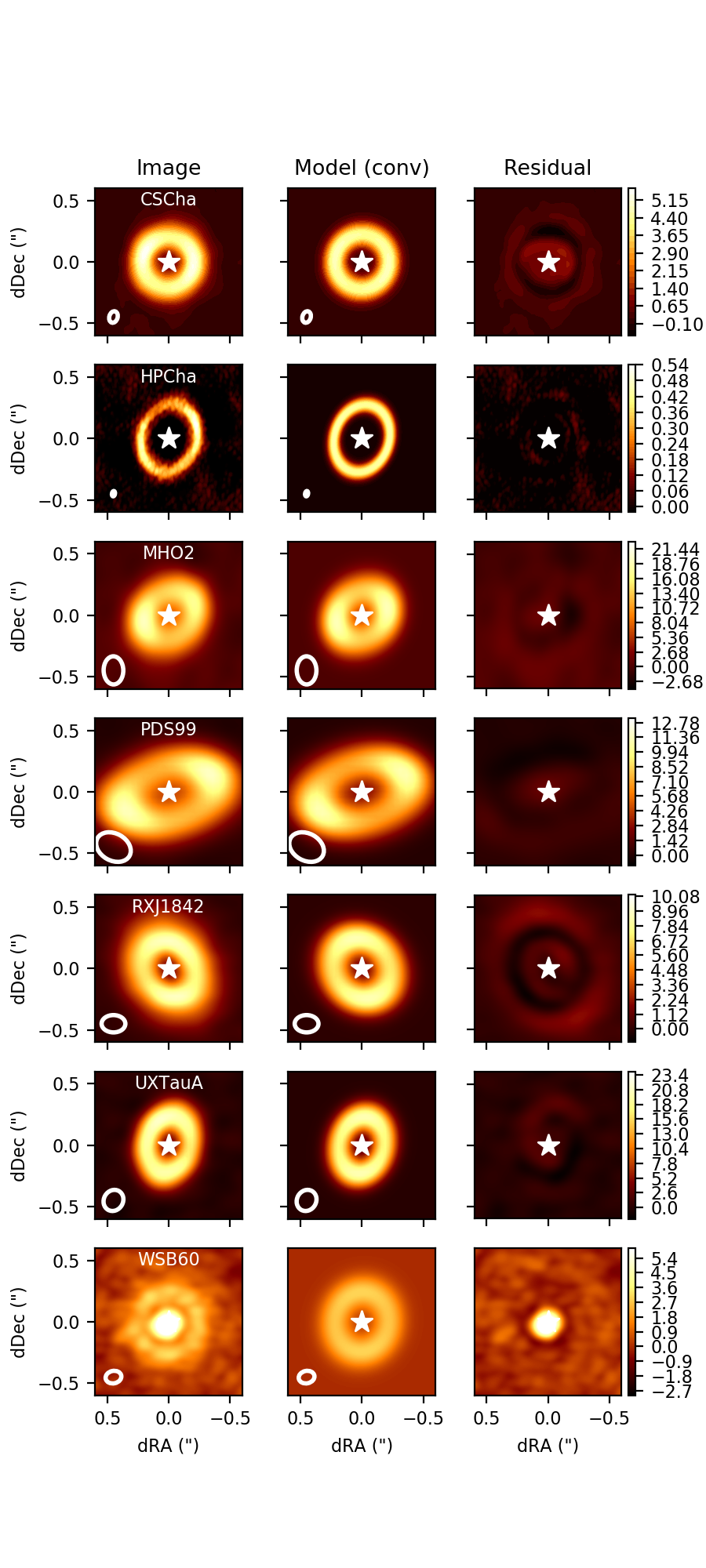}
    \caption{Best fit intensity models for the outer ring in 7 disks without previous information on the orientation. The model is a Gaussian ring and the best-fit parameters are given in Table \ref{tbl:outerdiskfit}. The white ellipse indicates the beam size.}
    \label{fig:outerdiskfit}
\end{figure}

Figure \ref{fig:outerdiskfit} shows the images and best-fit models. 

%\newpage
\section{Spectral energy distributions}

\begin{figure*}[htb]
    \centering
    \includegraphics[scale=1.0]{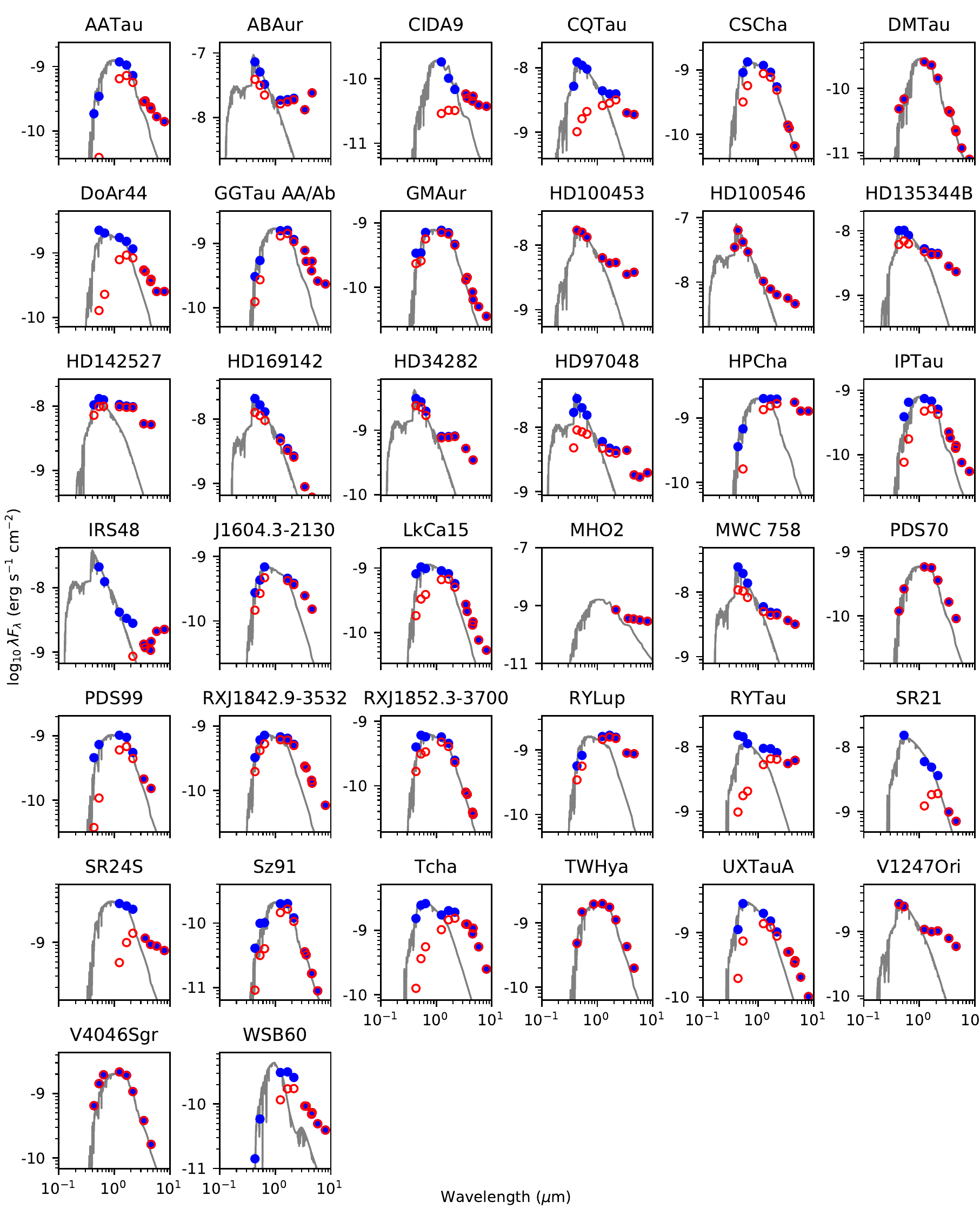}
    \caption{SEDs and stellar photosphere models for the 38 disks in our sample. The Kurucz photosphere models are shown with a gray line \citep{CastelliKurucz2004}. Red open circles indicate photometry before extinction correction; blue filled circles indicate corrected photometry.} 
    \label{fig:sed_gallery}
\end{figure*}




\listofchanges

\end{document}